\documentclass[lettersize,journal]{IEEEtran}
\usepackage{graphicx}
\usepackage{amssymb}
\usepackage{amsmath}
\usepackage{amsthm}
\usepackage{mathtools}
\newtheorem{theorem}{Theorem}
\newtheorem{assumption}{Assumption}

\usepackage{array,multirow,multicol,graphicx}
\usepackage{cite}
\usepackage{url} 
\usepackage{color}
\usepackage{float}
\usepackage{multirow}
\usepackage{subcaption}
\usepackage{array, tabularx, boldline}
\usepackage[ruled,lined,linesnumbered]{algorithm2e}
\usepackage{subfig}
\usepackage{graphicx}
\usepackage{subfig}
\usepackage{subcaption}
\usepackage[capitalize]{cleveref}
\Crefname{section}{Section}{Sections}
\Crefname{table}{Table}{Tables}
\usepackage{booktabs}
\usepackage{float}
\begin{document}
\title{Certifying the Right to be Forgotten: Primal-Dual Optimization for Sample and Label Unlearning in Vertical Federated Learning}

\author{Yu Jiang, Xindi Tong, Ziyao Liu, Xiaoxi Zhang, Kwok-Yan Lam, Chee Wei Tan % <-this % stops a space
\thanks{Yu Jiang and Kwok-Yan Lam are with the College of Computing and Data Science (CCDS), Nanyang Technological University, Singapore and Digital Trust Centre (DTC), Singapore.
Xindi Tong and Chee Wei Tan are with the College of Computing and Data Science (CCDS), Nanyang Technological University, Singapore. 
Ziyao Liu is with Digital Trust Centre (DTC), Singapore. 
Xiaoxi Zhang is with the School of Computer Science and Engineering, Sun Yat-sen University.
E-mail: yu012@e.ntu.edu.sg, to0002di@e.ntu.edu.sg, liuziyao@ntu.edu.sg, 
zhangxx89@mail.sysu.edu.cn, kwokyan.lam@ntu.edu.sg, cheewei.tan@ntu.edu.sg.
\textit{(Corresponding author: Chee Wei Tan.)}}

\thanks{Manuscript received April 19, 2005; revised August 26, 2015.}}

\markboth{Journal of \LaTeX\ Class Files,~Vol.~14, No.~8, August~2015}%
{Shell \MakeLowercase{\textit{et al.}}: Bare Demo of IEEEtran.cls for IEEE Journals}

\maketitle

\begin{abstract}
Federated unlearning has become an attractive approach to address privacy concerns in collaborative machine learning, for situations when sensitive data is remembered by AI models during the machine learning process. It enables the removal of specific data influences from trained models, aligning with the growing emphasis on the ``right to be forgotten." 
While extensively studied in horizontal federated learning, unlearning in vertical federated learning (VFL) remains challenging due to the distributed feature architecture. VFL unlearning includes sample unlearning that removes specific data points' influence and label unlearning that removes entire classes. Since different parties hold complementary features of the same samples, unlearning tasks require cross-party coordination, creating computational overhead and complexities from feature interdependencies.

To address such challenges, we propose FedORA (Federated Optimization for data Removal via primal-dual Algorithm), designed for sample and label unlearning in VFL. FedORA formulates the removal of certain samples or labels as a constrained optimization problem solved using a primal-dual framework. Our approach introduces a new unlearning loss function that promotes classification uncertainty rather than misclassification. An adaptive step size enhances stability, while an asymmetric batch design, considering the prior influence of the remaining data on the model, handles unlearning and retained data differently to efficiently reduce computational costs. We provide theoretical analysis proving that the model difference between FedORA and Train-from-scratch is bounded, establishing guarantees for unlearning effectiveness.
Experiments on tabular and image datasets demonstrate that FedORA achieves unlearning effectiveness and utility preservation comparable to Retrain with reduced computation and communication overhead.

\end{abstract}


\section{Introduction}
Federated learning (FL)~\cite{mcmahan2017communication,li2020federated,liang2024fedrema} enables collaborative model training without direct data sharing among multiple distributed parties. Nevertheless, AI models can remember sensitive data during the machine learning process, raising privacy concerns related to the ``right to be forgotten" established by regulations such as GDPR~\cite{regulation2018general}. Such regulations mandate not only the deletion of raw data upon user request but also the removal of its influence on trained models, necessitating solutions capable of removing specific data contributions from machine learning models while maintaining performance. In response, federated unlearning is an attractive approach to address privacy concerns in distributed environments by removing the influence of specific data points from trained models without requiring complete retraining~\cite{guo2019certified,liu2024guaranteeing,bourtoule2021machine, cao2015towards,jiang2024feduhb}.

To effectively implement federated unlearning, the underlying architecture needs to be considered. Federated learning can be classified into two categories based on data distribution patterns: horizontal federated learning (HFL)~\cite{liu2024privacy,boyd2011distributed,jiang2024efficient} where data is partitioned by samples, with each party holding identical features for different user subsets, and vertical federated learning (VFL)~\cite{liu2024vertical,yang2019federated,chen2020vafl, zhang2021survey} where data is partitioned by features, with multiple parties training a model on shared users with complementary feature subsets. While horizontal federated unlearning has received significant research attention, unlearning in VFL has garnered less focus~\cite{pan2025feature,gu2024few,han2025vertical} and presents greater challenges due to interparty dependencies, especially for sample and label unlearning.
In VFL systems, users may request removal of specific data samples to exercise data sovereignty, requiring sample unlearning. Additionally, regulatory policy changes may mandate withdrawal of entire data categories or classes, necessitating label unlearning to remove complete classes along with all associated samples.
Despite its importance, this problem remains largely unexplored. 
Removing samples or labels in VFL affects all participating parties simultaneously, as each party holds different feature subsets of the same samples. This process demands substantial computational and communication resources, while the interdependencies between distributed features complicate the isolation and removal of specific data contributions. Studies such as~\cite{varshney2025unlearning, gu2024few} have attempted to address this issue by adopting gradient ascent. However, existing methods often struggle to balance unlearning effectiveness and model performance preservation. Such approaches exhibit instability, sometimes failing to forget target data or causing excessive forgetting that degrades model utility~\cite{goodfellow2015empirical,KirkpatrickPRVD16}.

To address these challenges, we propose FedORA (Federated Optimization for data Removal via primal-dual Algorithm), adopting a primal-dual optimization framework for sample and label unlearning in VFL. 
FedORA formulates unlearning as a constrained optimization problem~\cite{beck2017first, smith2018cocoa, levy2020large}, inspired by Primal-Dual Hybrid Gradient (PDHG). 
The primal-dual framework naturally balances utility preservation on remaining data with effective forgetting of target data (unlearning data), while leveraging Lagrange duality to mathematically certify unlearning effectiveness.
Furthermore, we design an unlearning loss function that encourages uncertainty in classification for unlearning samples rather than pushing the model to misclassify unlearning data, thus avoiding excessive forgetting often associated with gradient ascent methods. In FedORA, stable convergence depends on appropriate step sizes for both primal and dual variable updates. We introduce an adaptive mechanism that tracks how much the model parameters change between successive iterations to dynamically adjust step sizes. 
Additionally, since the remaining data has already influenced the model during training, processing the entire dataset to maintain model utility during unlearning is unnecessary. FedORA employs an asymmetric batch processing strategy that processes only partial batches of remaining data and full batches of unlearning data, reducing computational overhead. Theoretically, we prove that the difference between the unlearned models obtained by FedORA and Train-from-scratch has an upper bound, providing theoretical validation of FedORA's unlearning effectiveness. Empirically, we evaluate the approach on the tabular dataset Income and image datasets MedMNIST, CIFAR-10, CIFAR-100, and Tiny-ImageNet across sample and label unlearning scenarios. Results indicate that FedORA achieves unlearning effectiveness comparable to Retrain with reduced computational and communication costs. Besides, we assess FedORA's resilience against membership inference attacks and backdoor attacks, validating that our approach effectively removes the influence of unlearning data.

\textbf{Our Contributions.}
The main contributions of this work are outlined below.
\begin{enumerate}
\item We propose FedORA, the first method for sample and label unlearning in VFL using a primal-dual optimization framework. By formulating unlearning as a constrained optimization problem, our method effectively balances unlearning effectiveness with model utility preservation.
\item We design a new unlearning loss function that encourages uncertainty in classification for unlearning data rather than misclassification. 
\item We introduce an adaptive step size mechanism that improves convergence stability, along with an asymmetric batch design that reduces computational costs.
\item We conduct comprehensive experimental analysis evaluating the trade-offs between unlearning effectiveness, model utility, and computational efficiency.

\end{enumerate}

\section{Related Work}
\subsection{Horizontal Federated Unlearning}
In HFL systems, the unlearning target may include either the complete removal of an entire client or specific subsets of data contributed by a client. To achieve unlearning across these varied targets, researchers have developed multiple methods, each offering different advantages depending on the specific implementation context~\cite{liu2024survey}.
Train-from-scratch (Retrain) rebuilds the model from scratch only using the remaining data after removing the target data~\cite{tao2024communication}. While Retrain ensures complete removal of target data contributions, it discards all previous training progress and requires significant computational and communication overhead, making it often impractical for large-scale federated systems. Gradient ascent applies reverse learning by increasing the loss on target data instead of reducing it, effectively reducing the influence of unlearning data~\cite{halimi2022federated}. However, gradient ascent can easily lead to forgetting too much information, requiring careful controls to preserve knowledge from remaining data. Fine-tuning calibrates the existing model using only the remaining data to reduce the impact of target data~\cite{jiang2024towards}. While more efficient than Retrain, fine-tuning still involves multiple iterations and increased storage costs. Model scrubbing applies mathematical changes to model parameters to approximate a model trained only on remaining data without actual retraining~\cite{cao2023fedrecover}. Most scrubbing approaches rely on mathematical approximations, which are computationally challenging for high-dimensional models. Synthetic data methods replace target data with synthetic data to help the model forget specific information~\cite{li2023federated, xu2023revocation}. For example, synthetic labels can be generated for target data and used during training to accomplish unlearning.

\begin{figure}
    \centering
    \includegraphics[width=\linewidth]{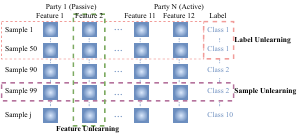}
    \caption{Difference among sample, label and feature unlearning.}
    \label{fig:unlearning}
\end{figure}
\subsection{Vertical Federated Unlearning}
\cref{fig:unlearning} illustrates the fundamental distinctions between three unlearning objectives within vertical federated learning: sample unlearning, label unlearning, and feature unlearning.
Sample unlearning removes partial samples from specific classes while preserving other samples. Label unlearning completely removes entire classes along with all associated samples. Feature unlearning addresses the unique VFL scenario by removing feature contributions from specific participating parties regardless of samples or classes. Research on VFU remains in its nascent stages. Deng et al.~\cite{deng2023vertical} modify model parameters through inverse gradient operations or parameter adjustments for unlearning. Wang et al.~\cite{wang2024efficient} propose to accelerate retraining using optimizer switching (RAdam to SGDM) and checkpoint loading. Pan et al.~\cite{pan2025feature} apply feature projection, masking, or transformation techniques to selectively forget feature-based information, while Han et al.~\cite{han2025vertical} use adversarial techniques to verify unlearning effectiveness. Varshney et al.~\cite{varshney2025unlearning} handle client-level unlearning through contribution isolation, while feature-level unlearning may involve feature masking or projection.
Current VFU research primarily focuses on feature unlearning, with limited attention to sample and label unlearning. Gu et al.~\cite{gu2024few} and Varshney et al.~\cite{varshney2025unlearning} employ gradient ascent to forget specific examples or labels, but suffer from instability issues. Sample and label unlearning remain crucial components within the VFL framework. To address this gap, our research introduces a primal-dual optimization framework that effectively balances unlearning effectiveness, computational efficiency, and model performance preservation.

\begin{figure}
    \centering
    \includegraphics[width=\linewidth]{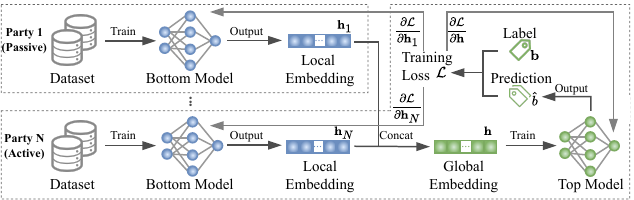}
    \caption{The VFL framework adopted in the paper.}
    \label{fig:vfl}
\end{figure}

\section{Preliminaries}
\subsection{Workflow of VFL}\label{sec:vfl}
VFL allows multiple parties, each holding distinct subsets of features for a shared user set, to collaboratively train a machine learning model while preserving data privacy. 
Let $\mathcal{P} = \{P_1, P_2, \ldots, P_N\}$ represent the set of $N$ parties, where each party $P_i$ maintains a local user (sample) ID set $\mathcal{U}_i$ and a feature matrix $\mathbf{A}_i \in \mathbb{R}^{|\mathcal{U}_i| \times d_i}$, with $d_i$ denoting the number of features possessed by $P_i$. Among these, the $N$-th party, which owns the labels, is designated as the active party, while the remaining parties function as passive parties. Each passive party holds its local data $\mathcal{D}_i$ and implements a trainable neural network serving as the bottom model $\theta_i$. The active party $P_N$ employs a trainable top model $\theta_q$ that integrates the outputs from all bottom models to generate predictions.

Prior to training, private set intersection (PSI)~\cite{pinkas2018scalable}, a cryptographic protocol, is employed to securely identify the common user set among all participating parties while ensuring that private information about non-overlapping users remains protected. This yields the intersection set $\mathcal{U} = \bigcap_{i=1}^N \mathcal{U}_i$, which contains only the overlapping user IDs. Each party then extracts the corresponding feature matrix \( \mathbf{A}_i \in \mathbb{R}^{|\mathcal{U}| \times d_i} \) to ensure consistent input for collaborative learning.

\subsubsection{Forward Propagation}
During the training process, each passive party \(P_i\) independently performs forward propagation using its bottom model. Specifically, it computes local embeddings
$\mathbf{h}_i = \phi_i(\mathbf{A}_i; \theta_i),$ where \(\phi_i: \mathbb{R}^{|\mathcal{U}| \times d_i} \to \mathbb{R}^{|\mathcal{U}| \times d_h}\) represents a learnable transformation function. These embeddings \(\mathbf{h}_i\) are then securely transmitted to the active party.
The active party \(P_N\) concatenates the embeddings to form the global embedding representation: $\mathbf{h} = \text{concat}(\mathbf{h}_1, \mathbf{h}_2, \ldots, \mathbf{h}_N)$. 
Subsequently, the active party uses this global embedding $\mathbf{h}$ to generate predictions through its top model $\theta_q$: $\hat{b} = u(\mathbf{h}; \theta_q)$, where $u$ is the output layer parameterized by $\theta_q$. For each sample $j$, the corresponding prediction is given by $\hat{b}_j = u(\mathbf{h}_j; \theta_q)$, where $\mathbf{h}_j$ denotes the global embedding of sample $j$. The active party, which holds the label vector $\mathbf{b} \in \mathbb{R}^{|\mathcal{U}|}$, computes the global loss function $\mathcal{L}(\Theta; \mathbf{A}, \mathbf{b}) = \frac{1}{|\mathcal{U}|} \sum_{j=1}^{|\mathcal{U}|} \psi \big(\hat{b}_j, b_j \big)$,
where $\Theta = \{\theta_1, \theta_2, \ldots, \theta_N, \theta_q\}$ includes all trainable parameters, $\mathbf{A} = [\mathbf{A}_1, \mathbf{A}_2, \ldots, \mathbf{A}_N]$ represents the concatenated feature matrix for the shared user set, and $\psi(\hat{b}_j, b_j)$ denotes a loss function measuring the discrepancy between prediction $\hat{b}_j$ and the ground truth label $b_j$.

\subsubsection{Backpropagation and Model Update}
Once the loss is computed, the active party derives the gradient $\frac{\partial \mathcal{L}}{\partial \mathbf{h}}$ with respect to the global embedding $\mathbf{h}$ and updates the top model parameters $\theta_q$ via standard backpropagation. Subsequently, the gradient information is decomposed and securely transmitted to each passive party $P_i$, allowing them to update their respective bottom models $\theta_i$. Specifically, the bottom model parameters are updated using gradient descent $\theta_i^{t+1} = \theta_i^t - \eta \frac{\partial \mathcal{L}}{\partial \theta_i}$,
where $\frac{\partial \mathcal{L}}{\partial \theta_i} = \frac{\partial \mathcal{L}}{\partial \mathbf{h}_i} \cdot \frac{\partial \mathbf{h}_i}{\partial \theta_i}$ and \( \eta > 0 \) is the learning rate.
The VFL framework adopted in this paper is shown in \cref{fig:vfl}.

\subsection{Notation}
We summarize the notations with descriptions used throughout the paper in \cref{tab:notations}.
\begin{table}[t!]
\centering
\caption{Summary of notations with description}
\label{tab:notations}
\begin{tabular}{cl}
\toprule
Notation & Description \\ \midrule
$i$ & Party index \\
$j$ & Sample index \\
$m$ & Number of unlearning classes \\
$n$ & Percentage of unlearning samples\\
$\mathcal{U}$& Sample set \\
$\gamma$ & Unlearning threshold \\
$\tau$ & Primal step size\\
$\sigma$ & Dual step size\\
$\delta$ & Batch selection ratio\\
$\omega$ & forgetting weight\\
$B$ & Batch size\\
$C$ & Number of classes \\
$\mathbf{b} $ & Label vector\\
$\theta_i$ & Bottom model parameters of party $P_i$ \\
$\theta_q$ & Top model parameters of party $P_N$ \\
$ \mathbf{A}_i$ &  Feature matrix of party $P_i$\\
$\mathcal{U}_i$ & Local user (sample) ID set hold by party \(P_i\)\\
$\mathcal{D}_u$ & Set of unlearning (target) samples \\
$\mathcal{D}_r$ & Set of remaining samples \\
$\mathbf{h}_j $  & Global embedding of sample $j$\\
 \( \psi(\mathbf{h}_j, b_j) \) & Loss function \\
$\mathbf{h}_i = \phi_i(\mathbf{A}_i; \theta_i)$ & Local embedding of party $P_i$\\
$\hat{b}_j = u(\mathbf{h}_j; \theta_q)$ & Prediction for sample \(j\)\\
\bottomrule
\end{tabular}
\end{table}

\begin{figure*}
    \centering
    \includegraphics[width=\linewidth]{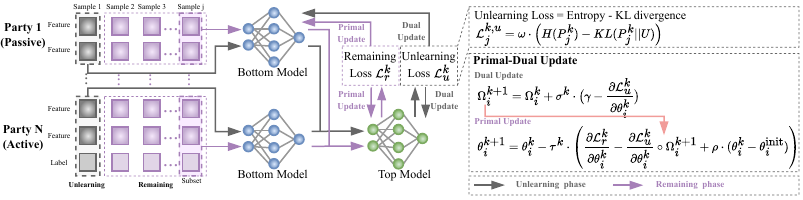}
    \caption{Overview of FedORA. Primal and dual variables are updated to achieve unlearning while maintaining model utility.}
    \label{fig:vfu}
\end{figure*}

\subsection{General Setting of PDHG algorithm}\label{sec:pdhg}
PDHG~\cite{sadiev2022communication,chambolle2016introduction,malitsky2018first}, also known as the Chambolle-Pock algorithm, has become a widely used optimization tool in machine learning applications. Initially introduced for solving convex-concave saddle-point problems~\cite{chambolle2011first}, PDHG has proven highly effective for handling optimization tasks involving non-smooth terms and linear operators. 
In its most general formulation, PDHG  addresses the following optimization problem:
\begin{equation}
    \min_{x \in \mathcal{X}} G(x) + F(Kx),
\end{equation}
where the problem is situated in the following mathematical framework: $x \in \mathcal{X}$ denotes the primal variable residing in a real Hilbert space $\mathcal{X}$, $G: \mathcal{X} \to \mathbb{R} \cup \{+\infty\}$ represents a proper, lower semicontinuous convex function, and $F: \mathcal{Y} \to \mathbb{R} \cup \{+\infty\}$ constitutes another proper, lower semicontinuous convex function defined on the dual space $\mathcal{Y}$. The operator $K: \mathcal{X} \to \mathcal{Y}$ serves as a bounded linear mapping that establishes the crucial coupling between the primal and dual spaces, characterized by its operator norm $\|K\|$.
Through the mathematical principle of duality, the primal optimization problem admits an equivalent saddle-point reformulation:
\begin{equation} \label{eq:saddle_point}
    \min_{x \in \mathcal{X}} \max_{y \in \mathcal{Y}} \langle Kx, y \rangle - F^*(y) + G(x), 
\end{equation}
where $y \in \mathcal{Y}$ represents the dual variable. The function $F^*(y)$ denotes the Fenchel-Rockafellar conjugate of $F(y)$, a fundamental concept in convex analysis, defined as:
\begin{equation}
    F^*(y) = \sup_{z \in \text{dom}(F)} \langle y, z \rangle - F(z).
\end{equation}
The dual problem corresponding to the saddle-point problem in \cref{eq:saddle_point} is derived from the primal problem through the concept of duality. The dual problem reformulates the original optimization problem by focusing on optimizing the dual variable $y$, and can be expressed as:
\begin{equation}
    \max_{y \in \mathcal{Y}} -F^*(y) - G^*(-K^\top y),
\end{equation}
where $G^*$ represents the Fenchel-Rockafellar conjugate of $G$, defined as:
\begin{equation}
    G^*(z) = \sup_{x \in \text{dom}(G)} \langle x, z \rangle - G(x),
\end{equation}
and $K^\top: \mathcal{Y} \to \mathcal{X}$ denotes the adjoint operator of $K$. The dual formulation naturally complements the primal problem, offering a perspective where the roles of $G$ and $F$ are interchanged via their conjugates.

\section{Methodology}
\subsection{Overview}
Existing VFL unlearning approaches primarily rely on gradient ascent that maximizes loss on target samples to achieve forgetting. However, such approaches often trigger catastrophic forgetting and requires careful regularization to preserve model performance, creating a challenging balance between effective unlearning and utility preservation.
To overcome these limitations, we propose FedORA, which formulates VFU as a constrained optimization problem solved through a primal-dual framework. Our key insight is that effective unlearning should promote classification uncertainty rather than misclassification, thereby avoiding the instability issues inherent in gradient ascent methods.
FedORA introduces several technical innovations: (i) a new uncertainty-based unlearning loss that encourages uniform probability distributions over target samples, (ii) an adaptive step size mechanism that ensures stable convergence, and (iii) an asymmetric batch processing strategy that reduces computation and communication overhead by leveraging the fact that remaining data has already influenced the model during training.
In the following parts, we detail the mathematical formulation of our primal-dual framework, present the design of FedORA, and provide theoretical analysis. The overview of FedORA is presented in \cref{fig:vfu}.

\subsection{Primal-dual Framework for Unlearning}
In this part, we transform the sample unlearning problem into a primal-dual optimization framework that efficiently handles constrained optimization. Unlearning aims to minimize loss on remaining data while ensuring target samples are effectively unlearned.
The primal-dual method preserves performance on remaining data during the primal stage. The dual variables track unlearning effectiveness by measuring how well the model has unlearned target samples. When the model still remembers target samples, dual variables increase to apply unlearning pressure. When forgetting is successful, dual variables decrease, automatically balancing utility preservation with effective unlearning.

Specifically, when the target users request unlearning, let \( \mathcal{D}_i^u \) denote the unlearning samples for each party \( i \), and \( \mathcal{D}_i^r \) denote the remaining samples for each party \( i \). 
For each participating party $P_i$, the local feature matrix $\mathbf{A}_i$ is partitioned into two distinct subsets:
\begin{equation}
\mathbf{A}_i = \begin{bmatrix} \mathbf{A}_i^r \\ \mathbf{A}_i^u \end{bmatrix},
\end{equation}
where $\mathbf{A}_i^u$ corresponds to the features of unlearning samples and $\mathbf{A}_i^r$ to the features of the remaining samples. 
Then, let $\mathcal{D}_u$ denote the set of samples (or user data points) designated for unlearning from all parties, denoted as \( \mathcal{D}_u = \bigcup_{i=1}^{N} \mathcal{D}_i^u \). Similarly, $\mathcal{D}_r$ represents the remaining samples from all parties, denoted as \( \mathcal{D}_r = \bigcup_{i=1}^{N} \mathcal{D}_i^r \). 
To formally define the unlearning objective, we introduce $\mathcal{L}_{u}(\Theta)$ as the loss function associated with the unlearning samples and $\mathcal{L}_{r}(\Theta)$ as the loss function for the remaining samples. The optimization problem for unlearning is formulated as:
\begin{equation}
\min_{\Theta} \mathcal{L}_r(\Theta) \quad 
\text{subject to} \quad \mathcal{L}_u(\Theta) \geq \gamma,
\end{equation}
where $\gamma$ represents the desired threshold for unlearning, specifically how much we want the model to ``forget" the target data. Larger values of $\gamma$ impose stricter unlearning requirements.
The goal is to minimize the loss on the remaining data \(\mathcal{L}_r(\Theta)\) while ensuring the loss on the unlearning data \(\mathcal{L}_u(\Theta)\) is above a threshold $\gamma$. To prevent catastrophic forgetting, we also introduce a proximal term to keep the model parameters close to their initial values. This leads to:
\begin{equation}
\min_{\Theta} \mathcal{L}_r(\Theta) + \frac{\rho}{2}\|\Theta - \Theta_{\text{init}}\|_2^2 \quad \text{subject to} \quad \mathcal{L}_u(\Theta) \geq \gamma,
\end{equation}
where $\rho > 0$ is the proximal regularization parameter. {Introducing a nonnegative dual variable $\Omega$ turns the constraint into an adaptive pressure: $\Omega$ naturally acts as a certificate of unlearning strength, growing when the unlearning requirement is violated and relaxing when it is satisfied.}
We then use the method of Lagrange multipliers to incorporate the constraint into the objective function:
\begin{equation}
\mathcal{L}(\Theta, \Omega) = \mathcal{L}_r(\Theta) + \frac{\rho}{2}\|\Theta - \Theta_{\text{init}}\|_2^2 + \Omega( \gamma - \mathcal{L}_u(\Theta)),
\end{equation}
{where \(\Omega \geq 0\) serves as the dual variable that controls the trade-off between retaining performance on remaining data and enforcing unlearning of target data, thereby providing certification of the constraint. At optimality, either $\mathcal{L}_u(\Theta^*)=\gamma$ with $\Omega^*\ge0$, or  $\mathcal{L}_u(\Theta^*)>\gamma$ with $\Omega^*=0$.} Then, we can define the dual function:
\begin{equation}
\begin{aligned}
q(\Omega) &= \inf_{\Theta} \mathcal{L}(\Theta, \Omega)  \\
&= \inf_{\Theta} \left[ \mathcal{L}_r(\Theta) + \Omega(\gamma - \mathcal{L}_u(\Theta)) + \frac{\rho}{2}\|\Theta - \Theta_{\text{init}}\|_2^2 \right],
\end{aligned}
\end{equation}
The dual function $q(\Omega)$, which captures the lowest achievable primal objective under a fixed dual penalty $\Omega$, leads to our dual problem:
\begin{equation}
\max_{\Omega \geq 0} q(\Omega).
\end{equation}
Maximizing the dual function over $\Omega \ge 0$ yields the best possible lower bound on the constrained primal objective and provides certification that the unlearning constraint is fulfilled.
While we could solve the problem by alternating between minimizing the Lagrangian with respect to $\Theta$ and maximizing it with respect to $\Omega$, we can directly characterize the solution as a saddle point of the Lagrangian. This reformulation allows us to employ efficient saddle-point algorithms tailored for distributed optimization. Specifically, we seek a point $(\Theta^*, \Omega^*)$ that satisfies:
\begin{equation}
\mathcal{L}(\Theta^*, \Omega) \leq \mathcal{L}(\Theta^*, \Omega^*) \leq \mathcal{L}(\Theta, \Omega^*), \quad \forall \Theta, \forall \Omega \geq 0.
\end{equation}
The inequality characterizes the saddle-point condition, where no unilateral change in $\Theta$ or $\Omega$ can improve the Lagrangian. Here, the Lagrange dual structure plays a dual role, guaranteeing the feasibility of unlearning and certifying the optimality of the solution.
This characterization leads us to the following saddle-point problem:
\begin{equation}
\begin{aligned}
& \min_{\Theta} \max_{\Omega \geq 0} \mathcal{L}(\Theta, \Omega) \\
= &\min_{\Theta} \max_{\Omega \geq 0} \left[ \mathcal{L}_r(\Theta) + \frac{\rho}{2}\|\Theta - \Theta_{\text{init}}\|_2^2 + \Omega(\gamma -\mathcal{L}_u(\Theta)) \right].
\end{aligned}
\end{equation}

Inspired by the standard PDHG in \cref{eq:saddle_point}, we first rewrite our problem:
\begin{equation}
\min_{\Theta} \max_{\Omega \geq 0} \langle K\Theta, \Omega \rangle + G(\Theta) - F^*(\Omega),
\end{equation}
where:
\begin{align}
K\Theta &= - \mathcal{L}_u(\Theta),\\
G(\Theta) &= \mathcal{L}_r(\Theta) + \frac{\rho}{2}\|\Theta - \Theta_{\text{init}}\|_2^2,\\
F^*(\Omega) &= -\gamma \Omega + \iota_{\Omega \geq 0}(\Omega),
\end{align}
with $\iota_{\Omega \geq 0}$ being the indicator function of the non-negative orthant. 
{Here, $K\Theta$ represents the linearized coupling from the unlearning constraint, denoting the per-iteration first-order approximation of $-\mathcal{L}_u$ at the current iterate, i.e., $\langle K\Theta, \Omega \rangle \coloneqq -\Omega \langle \nabla \mathcal{L}_u(\Theta^k), \Theta - \Theta^k \rangle$~\cite{Condatpdhglinear}, while we keep the compact notation $K\Theta$ for readability. 
Meanwhile, $G(\Theta)$ captures the regularized loss on the remaining data, and $F^*(\Omega)$ encodes the dual constraint along with the non-negativity indicator. The algorithm then proceeds iteratively with the following updates:}
\begin{align}
\Omega^{k+1} &= \arg\max_{\Omega \geq 0} \langle K\Theta^k, \Omega \rangle - F^*(\Omega) - \frac{1}{2\sigma}\|\Omega - \Omega^k\|_2^2,\\
\Theta^{k+1} &= \arg\min_{\Theta} \langle K\Theta, \Omega^{k+1} \rangle + G(\Theta) + \frac{1}{2\tau}\|\Theta - \Theta^k\|_2^2.
\end{align}
The dual update performs a projected gradient ascent on the forgetting constraint residual $\gamma - \mathcal{L}_u(\Theta^k)$. This explicit form updates the dual variable by accumulating constraint violation and projecting back to the non-negative orthant.
These updates can be computed explicitly as:
\begin{equation}\label{eq:Omage_k+1}
    \Omega^{k+1} = \Pi_{\Omega \geq 0}\left(\Omega^k + \sigma( \gamma -\mathcal{L}_u(\Theta^k))\right),
\end{equation}
where $\Pi_{\Omega \geq 0}$ represents projection onto the non-negative orthant. For \cref{eq:Omage_k+1}, we take the gradient of $\mathcal{L}_u(\Theta^k)$ to simplify the optimization problem with a first-order approximation so that the iterative updates can be calculated in the form of explicit gradient steps: 
\begin{align}
\Omega^{k+1} &= \Pi_{\Omega \geq 0}\left(\Omega^k + \sigma( \gamma\mathbf{1}- \nabla\mathcal{L}_u(\Theta^k) )\right).
\end{align}
The primal update minimizes a quadratic approximation of the objective, combining descent on the remaining data loss, repulsion from the unlearning data, and a proximal term centered at $\Theta_{\text{init}}$. This gradient-based update adjusts $\Theta$ to reduce $\mathcal{L}_r$ and avoid fitting the unlearning data, while softly anchoring to the initial parameters.
\begin{equation}
    \begin{aligned}
        \Theta^{k+1} =& \Theta^k - \tau (\nabla \mathcal{L}_r(\Theta^k) - \nabla \mathcal{L}_u(\Theta^k) \circ\Omega^{k+1} \\ 
        &+ \rho(\Theta^k - \Theta_{\text{init}})),
    \end{aligned}
\end{equation}
where $\circ$ denotes element-wise multiplication. The step sizes $\sigma$ and $\tau$ satisfy $\sigma\tau L^{2} < 1$, with $L$ denoting any global Lipschitz constant (or computable upper bound) of $\nabla\mathcal{L}_{u}(\Theta)$~\cite{Condatpdhglinear}.

\begin{algorithm}[t]
\fontsize{9.5pt}{11.4pt}\selectfont
\caption{FedORA }
\label{alg:FedORA-unlearning}
\SetAlgoLined
\KwIn{Unlearning sample set $\mathcal{D}_u$, remaining sample set $\mathcal{D}_r$, initial model parameters $\Theta^\text{init} $, initial dual step sizes $\sigma^0$, initial primal step sizes $\tau^0$, maximum dual step sizes $\sigma_{\max}$, maximum primal step sizes $\tau_{\max}$, unlearning threshold $\gamma$, forgetting weight $\omega$, number of classes $C$, uniform distribution $U$}
\KwOut{Unlearned model $\Theta^K$}
\For{$k = 0, 1, \ldots, K-1$}{
    \SetKwFunction{FU}{\textbf{Unlearning Phase}}
    \SetKwProg{Fn}{}{:}{}
    \Fn{\FU}{
    \For{each passive party $P_i$ \textbf{in parallel}}{
        Compute local embeddings $\mathbf{h}_i^{k,u} = \phi_i(\mathbf{A}_i^u; \theta_i^k)$\;
        Send $\mathbf{h}_i^{k,u}$ to the active party $P_N$\;
    }    
    \For{active party $P_N$ }{
    Compute local embeddings $\mathbf{h}_N^{k,u} = \phi_N(\mathbf{A}_N^u; \theta_N^k)$\;
    Form global embedding $\mathbf{h}_u^k = \text{concat}(\mathbf{h}_1^{k,u}, \mathbf{h}_2^{k,u}, \ldots, \mathbf{h}_N^{k,u})$\;
    Compute predictions $\hat{b}_u^k = u(\mathbf{h}_u^k; \theta_q^k)$\;    
    \For{each unlearning sample $j \in \mathcal{D}_u$}{
        Compute predicted probabilities $P_j^k = \text{softmax}(\hat{b}_j^{k,u})$\;
        Compute entropy $H(P_j^k) = -\sum_{c=1}^C P_{j,c}^k \log P_{j,c}^k$\;
        Compute KL divergence $KL(P_j^k || U) = \sum_{c=1}^C P_{j,c}^k \log \frac{P_{j,c}^k}{U_c}$\;
        Compute sample unlearning loss $\mathcal{L}_j^{k,u} = \omega \cdot \left(H(P_j^k) - KL(P_j^k || U)\right)$\;
    }   
    Compute unlearning loss $\mathcal{L}_u^k = \sum_{j \in \mathcal{D}_u} \mathcal{L}_j^{k,u}$\;
    Compute embedding gradients $\frac{\partial \mathcal{L}_u^k}{\partial \mathbf{h}_u^k}$\;
    Send gradients to each passive party\;
    }   
    \For{each party $P_i$ \textbf{in parallel}}{
        Compute parameter gradients $\frac{\partial \mathcal{L}_u^k}{\partial \theta_i^k}$\;
        Update dual variables $\Omega_i^{k+1} =\Pi_{\Omega \geq 0}\left( \Omega_i^k + \sigma^k \cdot \left( \gamma - \frac{\partial \mathcal{L}_u^k}{\partial \theta_i^k} \right)\right)$\;
    }
    }
    \SetKwFunction{FU}{\textbf{Remaining Phase}}
    \SetKwProg{Fn}{}{:}{}
    \Fn{\FU}{
    See in \cref{alg:FedORA-remaining}
    }
    $\tau^{k+1} = \min(\tau^k \cdot \xi \!\left(\frac{\Delta \Theta^{k+1}}{\Delta \Theta^{k}}\right), \tau_{\max})$\;
    $\sigma^{k+1} = \min(\sigma^k \cdot \xi \!\left(\frac{\Delta \Theta^{k+1}}{\Delta \Theta^{k}}\right), \sigma_{\max})$\;
}

\Return{Unlearned model $\Theta^K$}\;
\end{algorithm}

\subsection{Design of FedORA} \label{sec:FedORA}
Under the primal-dual framework, we propose FedORA for sample and label unlearning in VFL. The algorithm incorporates an uncertainty-based unlearning loss that encourages classification uncertainty rather than misclassification, an adaptive step size mechanism for stable convergence, and an asymmetric batch strategy that processes full batches for unlearning data while using mini-batches for the remaining data to reduce computational overhead.

\subsubsection{Iterative Optimization Phase}
FedORA operates through an iterative optimization process. The unlearning phase updates dual variables and the remaining phase updates primal variables, eliminating the influence on target data while preserving model performance on remaining data.

\textbf{Unlearning Phase (Dual Variable Update):}
During the $k$-th iteration of the unlearning process, for the unlearning sample set \(\mathcal{D}_i^u\), each passive party \(P_i\) begins by computing local embeddings using their respective models:
\begin{equation}
\mathbf{h}_i^{k,u} = \phi_i(\mathbf{A}_i^u; \theta_i^k).
\end{equation}
These embeddings are securely transmitted to the active party, ensuring privacy preservation. The active party concatenates the received embeddings to form the global representation: \(\mathbf{h}^k_u = \text{concat}(\mathbf{h}_1^{k,u}, \mathbf{h}_2^{k,u}, \dots, \mathbf{h}_N^{k,u})\), from which the predicted outputs: \(\hat{b}^k_u = u(\mathbf{h}^k_u; \theta_q^k)\) are derived.
The unlearning objective focuses on the unlearning samples \( \mathcal{D}_u\), through a specialized unlearning loss function:
\begin{equation}
\mathcal{L}_u^k = \sum_{j \in \mathcal{D}_u} \mathcal{L}^{k,u}_j,
\end{equation}
where \(\mathcal{L}^{k,u}_j\) represents the sample-specific unlearning loss, designed as described in \cref{sec: unlearning loss design}. 

The active party computes embedding gradients \(\frac{\partial \mathcal{L}_u^k}{\partial \mathbf{h}^k_u}\) and securely distributes them to all passive parties. Each passive party then calculates its respective parameter gradients \(\frac{\partial \mathcal{L}_u^k}{\partial \theta_i^k}\), which are used to update the dual variables:
\begin{equation}\label{eq:dual update}
\Omega_i^{k+1} = \Omega_i^k + \sigma^k \cdot \big (\gamma-\frac{\partial \mathcal{L}_u^k}{\partial \theta_i^k}  \big),
\end{equation}
where $\sigma^k$ represents the dual step size at the $k$-th iteration and $\Omega_i$ functions as a certificate of unlearning strength, accumulating gradient directions that indicate how parameters should be adjusted to minimize the model’s reliance on the unlearning samples. During parameter updates, these accumulated gradients systematically reduce or eliminate the model’s ability to recognize patterns in the target samples. The unlearning phase for FedORA is presented in \cref{alg:FedORA-unlearning}.

\textbf{Remaining Phase (Primal Variable Update):}
Following the dual update, the remaining phase ensures the model preserves performance on the remaining dataset \(\mathcal{D}_i^r\). During the $k$-th iteration of the unlearning process, each passive party $P_i$ computes local embeddings \(\mathbf{h}_i^{k,r} = \phi_i(\mathbf{A}_i^r; \theta_i^k)\). These embeddings are securely transmitted to the active party, which concatenates them to form the global representation $\mathbf{h}^k_r = \text{concat}(\mathbf{h}_1^{k,r}, \mathbf{h}_2^{k,r}, \dots, \mathbf{h}_N^{k,r})$.
The active party then generates predictions  
\(\hat{b}^k_r = u(\mathbf{h}^k_r; \theta_q^k)\).
The remaining objective is quantified through a standard cross-entropy loss over the remaining samples:
\begin{equation}
\mathcal{L}_r^k = \frac{1}{|\mathcal{U}_r|} \sum_{j=1}^{|\mathcal{U}_r|} \psi(\hat{b}_j^{k,r}, b_j^{k,r}),
\end{equation}
where \(\psi\) represents the loss function. The active party computes embedding gradients \(\frac{\partial \mathcal{L}^k_r}{\partial \mathbf{h}^k_r}\) and securely distributes them to all passive parties. Each party then calculates their respective parameter gradients \(\frac{\partial \mathcal{L}^k_r}{\partial \theta_i^k}\), which are used in the primal variable updates:
\begin{equation} 
    \theta_i^{k+1} = \theta_i^k - \tau^k \cdot \left(\frac{\partial \mathcal{L}_r^k}{\partial \theta_i^k} - \frac{\partial \mathcal{L}_u^k}{\partial \theta_i^k} \circ \Omega_i^{k+1}  + \rho \cdot (\theta_i^k - \theta_i^{\text{init}})\right),
\end{equation}
where $\tau^k$ represents the primal step size at the $k$-th iteration.
This update balances three competing objectives.
The gradient term $\frac{\partial \mathcal{L}_r^k}{\partial \theta_i^k}$ enhances model performance on remaining samples, while the dual variable term $\Omega_i^{k+1}$ applies accumulated gradient information to systematically diminish the influence of unlearning samples. Additionally, the proximal term $\rho \cdot (\theta_i^k - \theta_i^{\text{init}})$ maintains parameters in proximity to their initial values, thereby creating a regularization effect that mitigates catastrophic forgetting during the unlearning process. These updates are also applied to the global parameter \( \theta_q^k \) at the active party, following the same optimization principle. The remaining phase for FedORA is presented in \cref{alg:FedORA-remaining}.

The optimization proceeds iteratively until convergence criteria, such as parameter stability or loss reduction thresholds, are met. Through this process, the model progressively eliminates the influence of unlearning samples while preserving accuracy on the remaining data. The complete FedORA algorithm is outlined in \cref{alg:FedORA-unlearning}.

\subsubsection{Unlearning Loss Design} \label{sec: unlearning loss design}
{FedORA fundamentally departs from conventional gradient ascent methods, which often destabilize the model by forcing it to increase loss on target samples, potentially leading to catastrophic forgetting. We argue that genuine unlearning should not result in misclassification but rather in the model's inability to distinguish between classes.} From an information-theoretic perspective, true forgetting requires maximum uncertainty in predictions, where the model outputs uniform probability distributions over unlearning samples.
To formalize this principle, we define maximum uncertainty as the condition where the model outputs a uniform probability distribution across all possible classes, corresponding to maximum entropy in the prediction distribution. Given a predicted probability distribution $P$ for an unlearning sample, our objective is to maximize entropy \(H(P)\), thereby pushing the model toward maximum uncertainty, while minimizing the Kullback-Leibler (KL) divergence \(KL(P || U)\) between the predicted distribution and a uniform distribution, ensuring that the model does not retain structured biases from prior training. 
Assuming that there are $C$ classes, for each sample \(j \in \mathcal{D}_u\) targeted for unlearning, we define the unlearning loss as: 
\begin{equation}
\mathcal{L}^{k,u}_j = \omega \cdot \left(H(P_j^k) - KL(P_j^k || U)\right),
\end{equation}
where \(P_j^k = \text{softmax}(\hat{b}_j^{k,u})\) denotes the predicted probability distribution, \(H(P_j^k)= -\sum_{c=1}^C P_{j,c}^k \log P_{j,c}^k\) represents its entropy, \(KL(P_j^k || U)= \sum_{c=1}^C P_{j,c}^k \log \frac{P_{j,c}^k}{U_c}\) measures its divergence from the uniform distribution \(U\), and \(\omega\) is a forgetting weight parameter controlling the strength of unlearning. The entropy term encourages uncertainty in predictions, while the negative KL-divergence term actively drives the distribution toward uniformity. Together, these components ensure that the model achieves ``maximum forgetting" for the target samples, effectively removing their influence while maintaining overall model stability.

\begin{algorithm}[t]
\caption{Remaining Phase in FedORA}
\label{alg:FedORA-remaining}
\SetAlgoLined
\KwIn{Regularization term $\rho$, batch selection ratio $\delta$, batch size $B$}
\KwOut{Updated model parameters $\Theta^{k+1}$}
    \SetKwFunction{FR}{\textbf{Remaining Phase}}
    \SetKwProg{Fn}{}{:}{}
    \Fn{\FR}{
$R \gets \lceil \frac{\delta |\mathcal{D}_r|}{B}\rceil$\;
\For{$s = 0, 1, \ldots, R-1$}{
    Select batch $\mathcal{B}_r^s \subset \mathcal{D}_r$\;    
    \For{each passive party $P_i$ \textbf{in parallel}}{
        Compute local embeddings $\mathbf{h}_i^{k,s,r} = \phi_i(\mathbf{A}_i^r[\mathcal{B}_r^s]; \theta_i^{k,s})$\;
        Send $\mathbf{h}_i^{k,s,r}$ to the active party $P_N$\;
    }
    
    \For{active party $P_N$ \textbf{in parallel}}{
    Compute local embeddings $\mathbf{h}_N^{k,s,r} = \phi_N(\mathbf{A}_N^r[\mathcal{B}_r^s]; \theta_N^{k,s})$\;
    Form global embedding $\mathbf{h}_r^{k,s} = \text{concat}(\mathbf{h}_1^{k,s,r}, \mathbf{h}_2^{k,s,r}, \ldots, \mathbf{h}_N^{k,s,r})$\;
    Compute predictions $\hat{b}_r^{k,s} = u(\mathbf{h}_r^{k,s}; \theta_q^{k,s})$\;
    Compute loss for sample $j \in \mathcal{B}_r^s$ $\mathcal{L}_r^{k,s} = \frac{1}{|\mathcal{B}_r^s|} \sum_{j \in \mathcal{B}_r^s} \psi(\hat{b}_j^{k,s,r}, b_j)$\;
    Compute embedding gradients $\frac{\partial \mathcal{L}_r^{k,s}}{\partial \mathbf{h}_r^{k,s}}$\;
    Send gradients to each passive party\;
    Compute parameter gradients $\frac{\partial \mathcal{L}_r^{k,s}}{\partial \theta_q^{k,s}}$\;
    Update $\theta_q^{k,s+1} = \theta_q^{k,s} - \tau^k \cdot \left(\frac{\partial \mathcal{L}_r^{k,s}}{\partial \theta_q^{k,s}} - \frac{\partial \mathcal{L}_u^k}{\partial \theta_q^k} \circ \Omega_q^{k+1} + \rho \cdot (\theta_q^{k,s} - \theta_q^{\text{init}})\right)$\;
    }
    \For{each passive party $P_i$ \textbf{in parallel}}{
        Compute parameter gradients $\frac{\partial \mathcal{L}_r^{k,s}}{\partial \theta_i^{k,s}}$\;
        Update substep parameters $\theta_i^{k,s+1} = \theta_i^{k,s} - \tau^k \cdot \left(\frac{\partial \mathcal{L}_r^{k,s}}{\partial \theta_i^{k,s}} - \frac{\partial \mathcal{L}_u^k}{\partial \theta_i^k} \circ \Omega_i^{k+1} + \rho \cdot (\theta_i^{k,s} - \theta_i^{\text{init}})\right)$\;
    }    
}
}
\Return{$\Theta^{k+1} = \{\theta_1^{k+1},  \ldots, \theta_N^{k+1}, \theta_q^{k+1}\}$}\;
\end{algorithm}

\subsubsection{Adaptive Step Sizes} \label{sec: adaptive step sizes}
In FedORA, stable convergence critically depends on appropriate step sizes for both primal and dual variable updates. We introduce an adaptive mechanism that dynamically adjusts these step sizes based on the relative change in parameter magnitudes between consecutive iterations. A central idea for adapting the step sizes is to monitor the norm of the difference between successive iterates. For the primal variables, we denote the change between iterations as
\begin{equation}
\Delta \Theta^k = \|\Theta^k - \Theta^{k-1}\|_2.
\end{equation}
A relatively small \(\Delta \Theta^k\) compared to the previous change \(\Delta \Theta^{k-1}\) suggests that the updates are minor and that a larger step size might safely speed up convergence. Conversely, if \(\Delta \Theta^k\) is significantly larger than \(\Delta \Theta^{k-1}\), this indicates potential instability, justifying a reduction in the step size for the subsequent iteration.
Using this metric, we adaptively adjust both the primal step size $\tau$ and dual step size $\sigma$:
\begin{align}
    \tau^{k+1} = \tau^{k} \cdot \xi \!\left(\frac{\Delta \Theta^{k+1}}{\Delta \Theta^{k}}\right), \\
    \sigma^{k+1} = \sigma^{k} \cdot \xi \!\left(\frac{\Delta \Theta^{k+1}}{\Delta \Theta^{k}}\right),
\end{align}
where the adjustment function $\xi(x)$ is defined as:
\begin{equation}
\xi (x) = 
\begin{cases}
\kappa_i & \text{if } x < \beta, \\
\kappa_d & \text{if } x > \alpha, \\
1 & \text{otherwise},
\end{cases}
\end{equation}
where $\alpha$ and $\beta$ denote threshold parameters satisfying $\beta < \alpha$, while $\kappa_d < 1$ and $\kappa_i > 1$ represent the decay and increase coefficients, respectively. To ensure stability, we further impose upper bounds on the step sizes:
\begin{equation}
\tau^{k+1} = \min(\tau^{k+1}, \tau_{\max}),~ \sigma^{k+1} = \min(\sigma^{k+1}, \sigma_{\max}).
\end{equation}

\subsubsection{Asymmetric Batch Design} \label{sec: asymmetric batch}
FedORA implements an asymmetric batch design that strategically differentiates between unlearning and preservation objectives. Since the unlearning phase aims to completely remove target samples from the model, it is necessary to process the entire unlearning dataset \(\mathcal{D}_u\) as a full batch. In contrast, the remaining dataset \(\mathcal{D}_r\) has already influenced the model during initial training, making full-batch training unnecessary. {Consequently, rather than processing the entire \(\mathcal{D}_r\) as a full batch that would incur substantial computational and communication costs, we select $\delta$ (\%) from \(\mathcal{D}_r\). This allows the model to continue adapting to the remaining data more efficiently, focusing computational resources on fine-tuning rather than complete retraining.}
We introduce an internal substep counter \(s\) within each main iteration \(k\). At the beginning of the remaining phase, we initialize the substep parameters with the current main iteration parameters $\theta_i^{k,0} = \theta_i^k$.
Then, we perform \(R\) consecutive parameter updates, each using a batch \(\mathcal{B}_r^s\) drawn from \(\mathcal{D}_r\). For each substep \(s\), the parameters evolve as follows:
\begin{equation}\label{eq:primal update}
\begin{aligned}
    \theta_i^{k,s+1} = & \theta_i^{k,s} - \tau^k \cdot (\frac{1}{|\mathcal{B}_r^s|} \sum_{j \in \mathcal{B}_r^s} \frac{\partial \mathcal{L}_r^{k}}{\partial \theta_i^{k,s}}\\
     & - \frac{\partial \mathcal{L}_u^k}{\partial \theta_i^k} \circ \Omega_i^{k+1} + \rho \cdot (\theta_i^{k,s} - \theta_i^{\text{init}})),
\end{aligned}
\end{equation}
where $R = \lceil \frac{\delta |\mathcal{D}_r|}{B}\rceil$ and $B = |\mathcal{B}_r^s|$ represents the batch size.
After completing all substeps, the final update for the main iteration is set as $\theta_i^{k+1} = \theta_i^{k,R}$.
{By processing only portions of \(\mathcal{D}_r\), our method achieves a balance between performance and efficiency in the vertical federated unlearning.}

\begin{table}[t]
\centering
\caption{{Computational and communication overhead}}
\label{tab:complexity_comparison}
\begin{tabular}{lcc}
\toprule
Method & Computation Complexity & Communication Complexity \\
\midrule
Retrain & $O( |D_r|K d)$ & $O( \frac{|D_r|}{B}K d_h)$ \\
GA & $O(|D_u|K d)$ & $O( \frac{|D_u|}{B}K d_h)$ \\
ICO & $O(|D_u| d +   |D_r| \nu K d)$ & $O(\frac{|D_u|}{B} d_h + \frac{|D_r|}{B} \nu K d_h)$ \\
CVFU & $O((|D_r| + \upsilon) Kd)$ & $O( (\frac{|D_r|}{B} + \upsilon) Kd_h)$\\
FedORA & $O( (|D_u| + \delta |D_r|) Kd)$ & $O( (\frac{|D_u|}{B} + \delta \frac{|D_r|}{B}) Kd_h)$ \\
\bottomrule
\end{tabular}
\end{table}

\begin{table}[t]
\centering
\caption{Running time (sec) of each unlearning round}
\label{tab:running time}
\begin{tabular}{lccccc}
\toprule
         & Retrain & GA   & ICO & CVFU & FedORA\\ \midrule
Income    & 1.43    & 0.79   & 1.50   & 1.18   & 1.22 \\         
MedMNIST  & 9.09    & 1.30   & 4.34   & 4.08   & 2.86 \\
CIFAR-10  & 5.35    & 1.85   & 4.86   & 5.57   & 2.48 \\
CIFAR-100 & 13.12   & 4.14   & 6.59   & 13.21  & 5.84 \\
Tiny-ImageNet & 62.24 & 22.33  & 45.83 & 55.98 & 23.50 \\ \bottomrule
\end{tabular}
\end{table}

\begin{table*}[t]
\centering
\caption{{Test accuracy comparison across different sample unlearning settings}}
\centering
\label{tab:ta_sample}
\begin{tabular*}{\textwidth}{@{\extracolsep{\fill}}lcccccccccc}
\toprule
& \multicolumn{2}{c}{Income} & \multicolumn{2}{c}{MedMNIST} & \multicolumn{2}{c}{CIFAR-10} & \multicolumn{2}{c}{CIFAR-100} & \multicolumn{2}{c}{Tiny-ImageNet} \\
\cmidrule(lr){2-3} \cmidrule(lr){4-5} \cmidrule(lr){6-7} \cmidrule(lr){8-9} \cmidrule(lr){10-11}
& m=1, & m=2, & m=2, & m=5, & m=2, & m=5, & m=20, & m=50, & m=40, & m=100, \\
& n=50\% & n=25\% & n=50\% & n=50\% & n=50\% & n=50\% & n=50\% & n=50\% & n=50\% & n=50\% \\
\midrule
Retrain & 0.8291 & 0.8584 & 0.8542 & 0.8482 & 0.8155 & 0.8141 & 0.6572 & 0.6141 & 0.6259 & 0.5849 \\
GA & 0.7936 & 0.7891 & 0.8385 & 0.7948 & 0.6841 & 0.6932 & 0.6074 & 0.5828 & 0.5784 & 0.5551 \\
ICO & 0.8092 & 0.8098 & 0.8442 & 0.8241 & 0.8079 & 0.7877 & 0.6167 & 0.6086 & 0.5874 & 0.5797 \\
CVFU & 0.8284 & 0.8118 & 0.8547 & 0.8302 & 0.7848 & 0.7903 & 0.5311 & 0.5470 & 0.5058 & 0.5209 \\
FedORA & 0.8260 & 0.8548 & 0.8663 & 0.8589 & 0.8165 & 0.8156 & 0.6398 & 0.6145 & 0.6093 & 0.5853 \\
\bottomrule
\end{tabular*}
\end{table*}

\begin{table*}[t]
\centering
\caption{{Test accuracy comparison for different label unlearning settings}}
\label{tab:ta_label}
\begin{tabular*}{\textwidth}{@{\extracolsep{\fill}}lcccccccc}
\toprule
& \multicolumn{2}{c}{MedMNIST} & \multicolumn{2}{c}{CIFAR-10} & \multicolumn{2}{c}{CIFAR-100} & \multicolumn{2}{c}{Tiny-ImageNet} \\
\cmidrule(lr){2-3} \cmidrule(lr){4-5} \cmidrule(lr){6-7} \cmidrule(lr){8-9}
& m=2, & m=5, & m=2, & m=5, & m=20, & m=50, & m=40, & m=100, \\
& n=100\% & n=100\% & n=100\% & n=100\% & n=100\% & n=100\% & n=100\% & n=100\% \\
\midrule
Retrain & 0.8454 & 0.8175 & 0.8135 & 0.8065 & 0.6176 & 0.5905 & 0.5826 & 0.5571 \\
GA & 0.8189 & 0.7415 & 0.6745 & 0.6625 & 0.5862 & 0.5060 & 0.5530 & 0.4774 \\
ICO & 0.8325 & 0.7744 & 0.7839 & 0.7783 & 0.6056 & 0.5796 & 0.5713 & 0.5468 \\
CVFU & 0.8478 & 0.7330 & 0.7777 & 0.7905 & 0.5828 & 0.5425 & 0.5498 & 0.5118 \\
FedORA & 0.8025 & 0.7833 & 0.8132 & 0.8047 & 0.6317 & 0.5953 & 0.5960 & 0.5616 \\
\bottomrule
\end{tabular*}
\end{table*}

\subsection{Theoretic Analysis}
In this section, we establish a theoretical framework to analyze the difference between the model obtained by FedORA and the model obtained by Train-from-scratch. We first outline the assumptions underlying our theoretical analysis. Then, we demonstrate that the difference between the model obtained by FedORA and the model obtained through the Train-from-scratch method is bounded. A detailed proof is provided in Appendix \ref{Proof of Theorem}.

\begin{assumption} \label{assump:convex}
    The loss function $\mathcal{L}_r(\Theta)$ is $\mu$-strongly convex and $L$-smooth: 
   \begin{align}
        \langle \nabla \mathcal{L}_r(\Theta_1) - \nabla \mathcal{L}_r(\Theta_2), \Theta_1 - \Theta_2 \rangle \geq \mu \Vert \Theta_1 - \Theta_2\Vert_2^2, \\
         \Vert \nabla \mathcal{L}_r(\Theta_1) - \nabla \mathcal{L}_r(\Theta_2) \Vert_2 \leq L \Vert \Theta_1 - \Theta_2\Vert_2.
   \end{align}
    where $\mu, L > 0$.
\end{assumption}

\begin{assumption} \label{assump:minibatch}
    The mini-batch gradient estimation error in FedORA is bounded~\cite{bottou2018optimization}: 
    \begin{equation}
        \Vert \frac{1}{|\mathcal{B}_r|}\sum_{j \in \mathcal{B}_r} \nabla \mathcal{L}_r(\Theta; j) - \nabla \mathcal{L}_r(\Theta)\Vert_2 \leq \frac{\sigma_r}{\sqrt{|\mathcal{B}_r|}},
    \end{equation}
    where $\mathcal{B}_r$ denotes a mini-batch of remaining data and $\sigma_r^2 = \mathbb{E}[\Vert \nabla \mathcal{L}_r(\Theta; j) - \nabla \mathcal{L}_r(\Theta)\Vert_2^2]$ is the variance of the stochastic gradient on remaining data.
\end{assumption}

\begin{assumption} \label{assump:gradient}
    The gradient of the unlearning loss function $\mathcal{L}_u(\Theta)$ is $G$-bounded for all parameters $\Theta$ in the feasible region~\cite{beck2017first}: 
    \begin{equation}
        \Vert\nabla \mathcal{L}_u(\Theta)\Vert_2\leq G,
    \end{equation}
    where $G > 0$ is a constant that bounds the magnitude of the gradient norm throughout the optimization process.
\end{assumption}

\begin{assumption} \label{assump:dual_bound}
    The dual variable sequence $\{\Omega^k\}$ generated by the primal-dual algorithm is uniformly bounded: $\|\Omega^k\|_2 \leq \Omega_{max}$ for the constant $\Omega_{max} > 0$ and all $k \geq 0$~\cite{chambolle2011first}.
\end{assumption}

\begin{theorem}[Model difference upper bound between FedORA and Train-from-scratch]\label{theo}
\begin{equation}
\begin{aligned}
        \|\Theta^k -\bar{\Theta}^k\|_2 \leq & (\sqrt{1 - \tau \mu})^k \|\Theta^0 - \bar{\Theta}^0\|_2 \\
       & + \frac{\sqrt{\tau}}{1 - \sqrt{1-\tau\mu}} \left(\frac{\sigma_r}{\sqrt{|\mathcal{B}_r|}} + G \Omega_{max}\right),
\end{aligned}
\end{equation}
where $\Theta^k$ denotes the model parameters of FedORA and $\bar{\Theta}^k$ denotes the model parameters of Train-from-scratch.
\end{theorem}

Train-from-scratch refers to the approach of removing the data to be unlearned and training a new global model from scratch using only the remaining data. Therefore, we use Train-from-scratch as a benchmark to measure the unlearning effectiveness of our method, FedORA.
Based on Theorem \ref{theo}, we establish an upper bound on the difference between the unlearned model obtained by Train-from-scratch and FedORA, demonstrating that the difference between FedORA's solution and the ideal retrained solution is bounded. For the term $(\sqrt{1 - \tau \mu})^k \|\Theta^0 - \bar{\Theta}^0\|_2$, since $\tau \mu > 0$, we have $\sqrt{1 - \tau \mu} < 1$, which means this term decays exponentially with the number of iterations $k$. When $k \to \infty$, this term approaches zero. As $k \to \infty$, $    \lim_{k \to \infty} \|\Theta^k - \bar{\Theta}^k\|_2 \leq \frac{\sqrt{\tau}}{1 - \sqrt{1-\tau\mu}} \left(\frac{\sigma_r}{\sqrt{|\mathcal{B}_r|}} + G \Omega_{max}\right).$
The mini-batch noise will affect the difference between the two methods. Increasing the batch size $|\mathcal{B}_r|$ will reduce this difference, as larger batches provide more accurate gradient estimates and thus better approximation to the Train-from-scratch solution.

\begin{figure}[t]
    \centering
    \includegraphics[width=\linewidth]{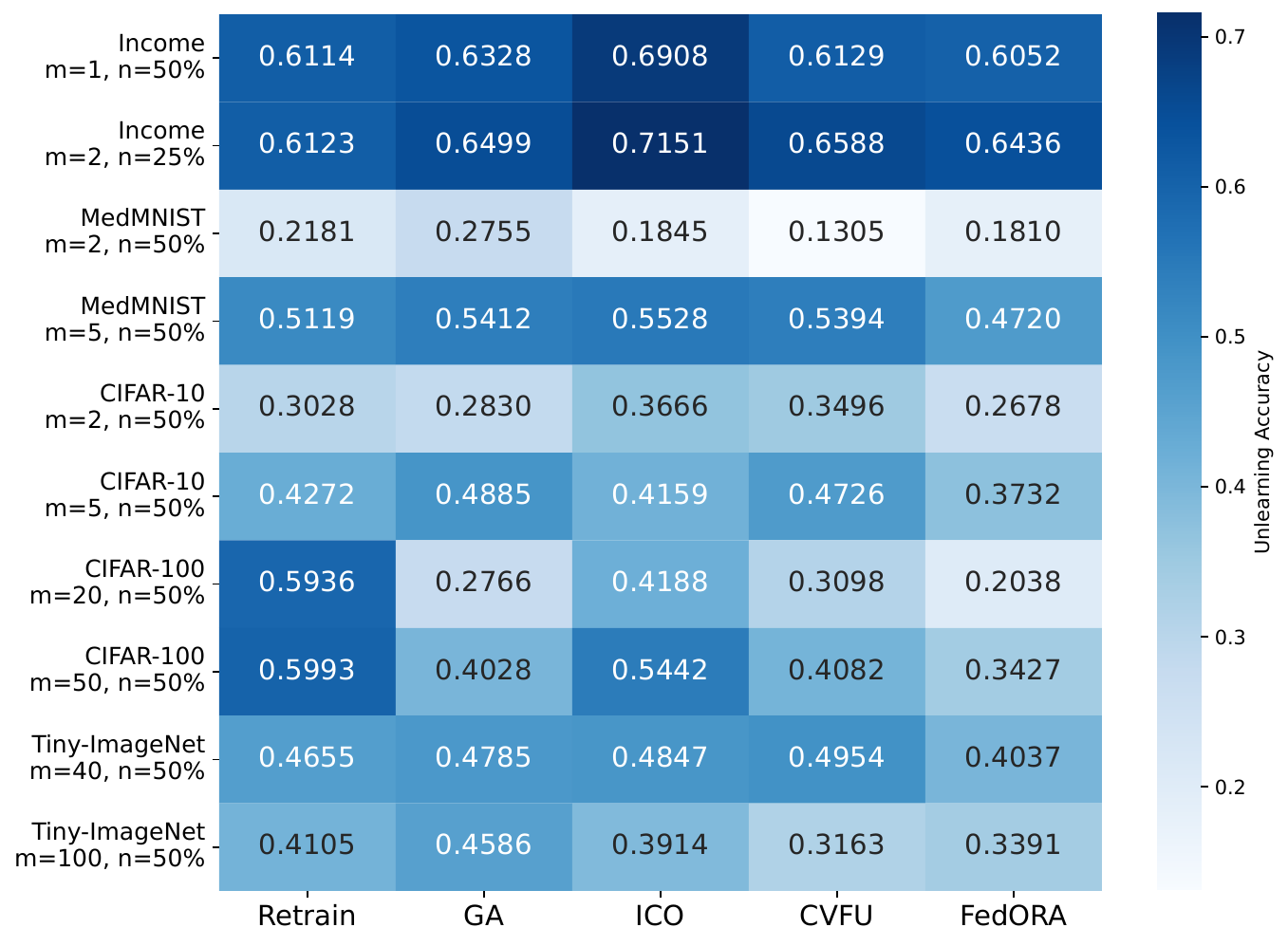}
    \caption{{Unlearning accuracy for different sample unlearning.}}
    \label{fig:ua_sample}
\end{figure}

\section{Experiments}
\subsection{Setup}
\subsubsection{Unlearning Pipeline in VFL} 
We establish a well-trained model by partitioning a neural network between passive parties (bottom models) and one active party (top model). Upon receiving sample deletion or label unlearning requests, all parties simultaneously remove the target samples. The system then transitions to the unlearning phase, aiming to eliminate the influence of specific data while preserving overall model utility. During the unlearning phase, we implement FedORA or other existing methods.
\subsubsection{Datasets and Models}
We evaluate our methods on five datasets with tailored VFL architectures.
\textbf{Income}~\cite{kohavi1996scaling} is a tabular dataset of 48,842 samples for binary income classification. We use VFL-based MLPs.
\textbf{MedMNIST}~\cite{yang2023medmnist} contains biomedical images across 9 tissue types. We use PathMNIST, a pathology dataset with 107,180 histopathology images of $28 \times 28$ pixels with 9 classes. We adopt a VFL-based ResNet18~\cite{he2016deep} architecture.
\textbf{CIFAR-10}~\cite{krizhevsky2009learning} has 60,000 RGB images of $32 \times 32$ pixels across 10 classes. We adopt a VFL-based ResNet18~\cite{he2016deep} architecture.
\textbf{CIFAR-100}~\cite{krizhevsky2009learning} extends CIFAR-10 with 100 classes. We use a VFL-based DenseNet~\cite{huang2017densely} architecture.
\textbf{Tiny-ImageNet}~\cite{le2015tiny} contains 200 classes with 100,000 training images at $64 \times 64$ pixels. We use a VFL-based ResNet50 architecture.

\subsubsection{Implementation Details} We simulate a VFL environment with two passive parties and one active party. 
For unlearning experiments, we conduct both sample and label unlearning on MedMNIST, CIFAR-10, CIFAR-100, and Tiny-ImageNet. Since Income is a binary classification dataset, we only perform sample unlearning. We configure different unlearning settings by selecting $m$ unlearning classes and removing $n$ samples, where $m$ represents the number of unlearning classes and $n$ represents the percentage of samples to unlearn within each unlearning class.

\subsubsection{Baseline}
We compare our proposed approach with four unlearning methods.
\textbf{Retrain} serves as the standard for unlearning, where the model is retrained using only the remaining data after removing the target samples to be forgotten. 
\textbf{Gradient ascent (GA)}~\cite{varshney2025unlearning, gu2024few} realizes sample unlearning in VFL by maximizing the loss on unlearning data.
{\textbf{ICO}~\cite{li2025inverse} computes the contribution of data to be forgotten to model parameters, then subtracts these contributions from current parameters and performs gradient correction optimization.
\textbf{CVFU}~\cite{wang2025forgetting} achieves asynchronous data unlearning by maintaining a confidence matrix at the active party and updating the confidence values within it.}

\begin{figure}[t]
    \centering
    \includegraphics[width=\linewidth]{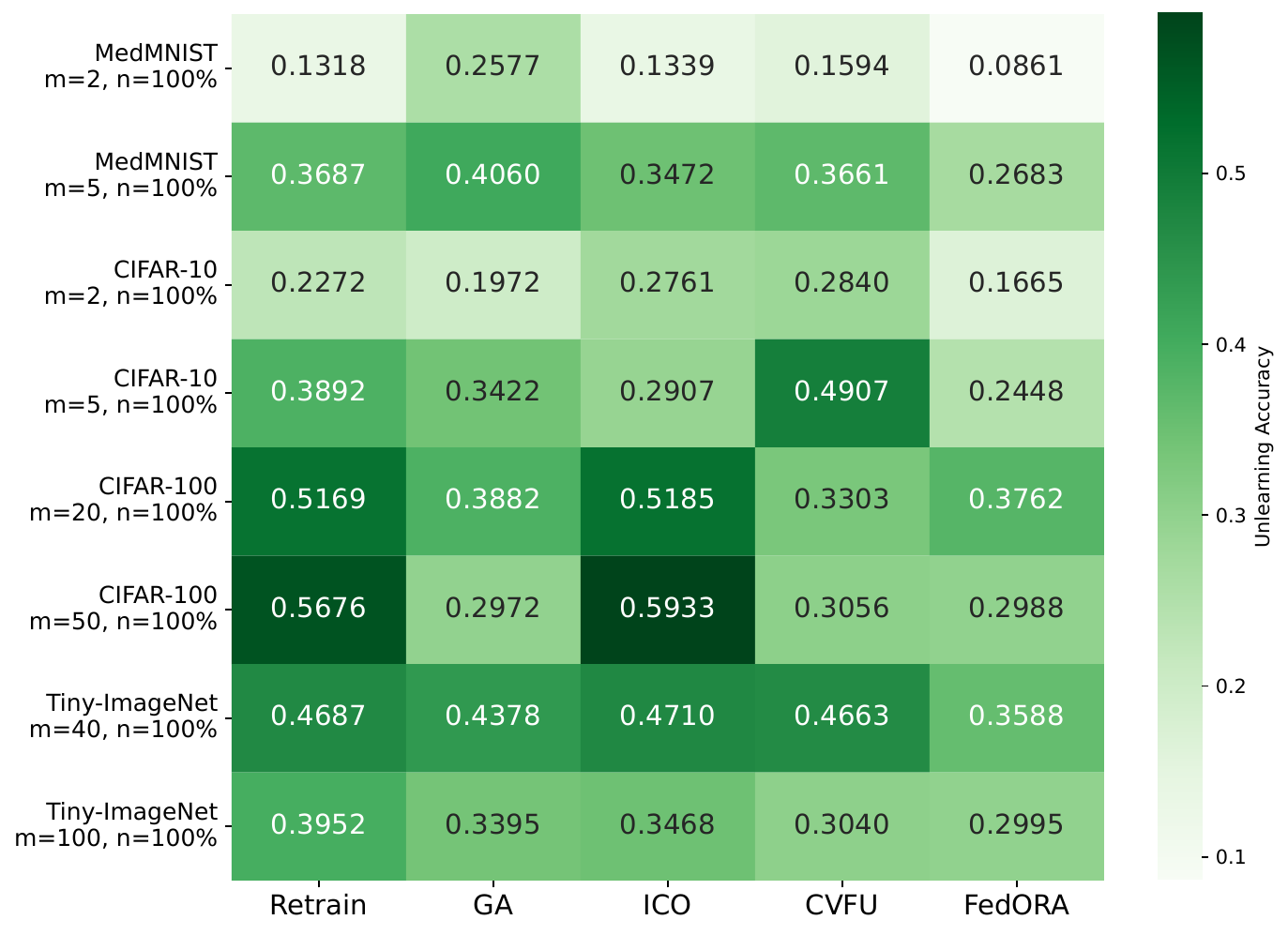}
    \caption{{Unlearning accuracy for different label unlearning.}}
    \label{fig:ua_label}
\end{figure}

\subsubsection{Metrics}  
We evaluate unlearning methods across three aspects: utility preservation, unlearning effectiveness, and attack resilience. We measure utility preservation using accuracy (Acc.) on the test data after unlearning, where higher test accuracy indicates better utility preservation. We evaluate unlearning effectiveness by measuring the accuracy on the unlearning dataset after unlearning, where lower unlearning accuracy indicates more effective forgetting, as the model should no longer recognize the samples it was supposed to forget.
Additionally, we employ membership inference attacks (MIA)~\cite{shokri2017membership} and backdoor attacks~\cite{liu2020reflection} to evaluate whether unlearning samples have been effectively removed. For MIA, we train a shadow model on the original model's outputs to distinguish training members from non-members. After unlearning, the shadow model evaluates whether the unlearning samples still exhibit membership patterns. A membership inference attack success rate (MIA-ASR) approaching 50\% indicates effective unlearning, demonstrating that the shadow model can no longer distinguish the unlearning samples from genuinely unseen data. For backdoor attacks, we inject a white-pixel trigger at the bottom-right corner of images into the unlearning samples before the original training phase, with a specific label as the target class. These triggered samples are trained to be classified as the target label during training. After unlearning, we test whether the unlearned model still responds to the backdoor triggers by measuring how many triggered samples are still classified as the target label. A low backdoor attack success rate (BD-ASR) indicates that the unlearning process has successfully removed the malicious trigger responses learned from the unlearning samples.

\begin{figure*}[t]
\centering
\begin{subfigure}[b]{0.19\textwidth}
    \includegraphics[width=\textwidth]{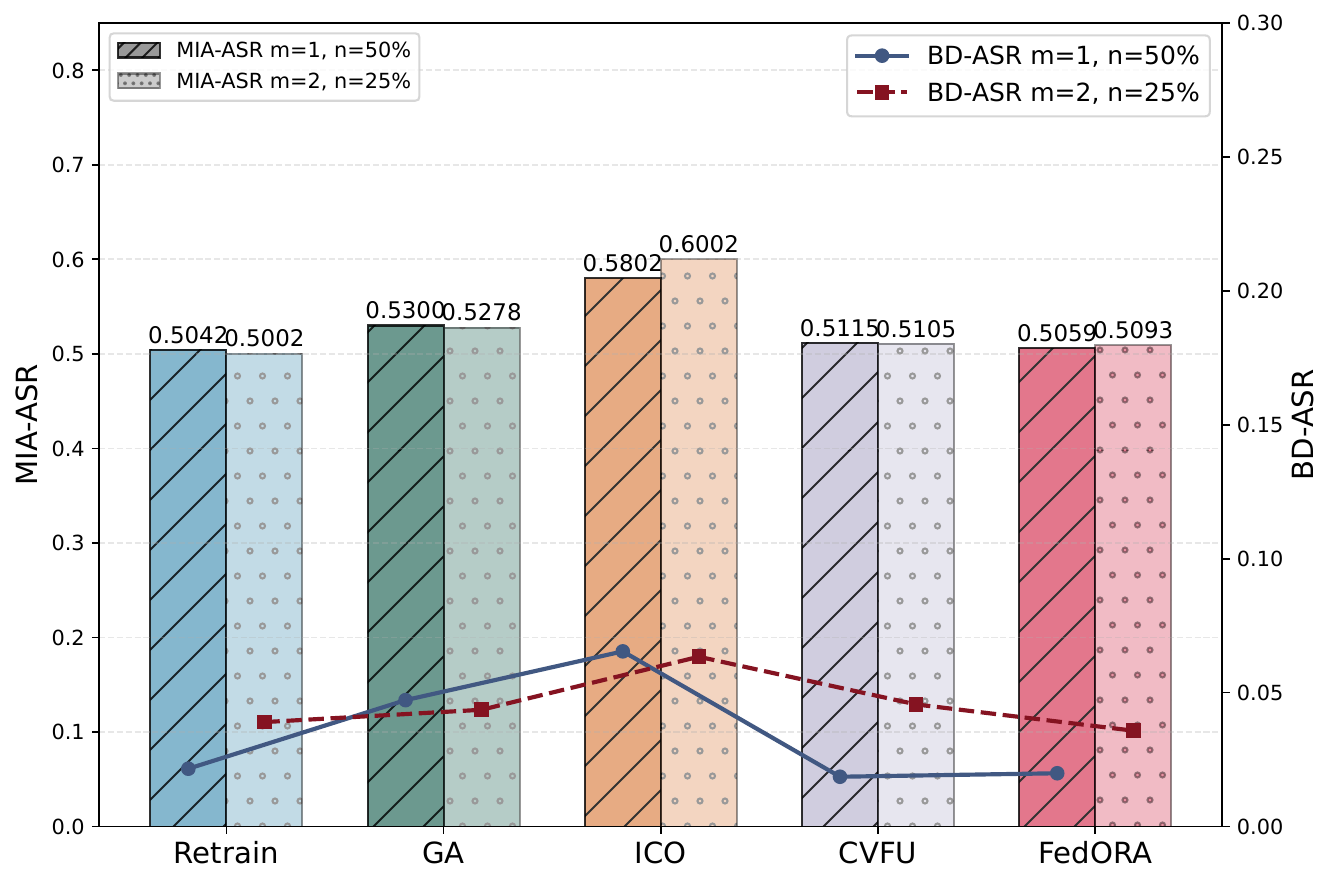}
    \caption{Income}
\end{subfigure}
\begin{subfigure}[b]{0.19\textwidth}
    \includegraphics[width=\textwidth]{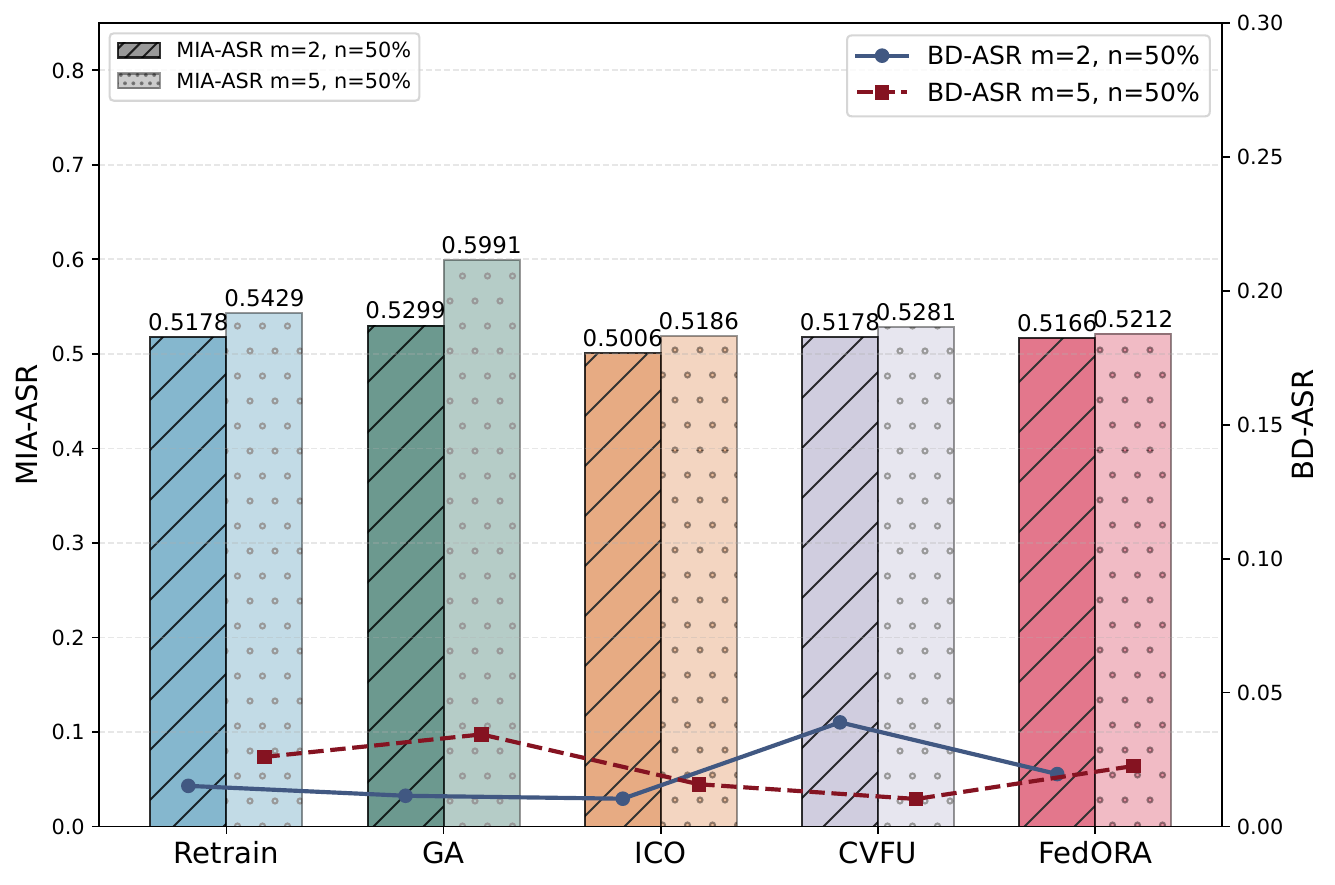}
    \caption{MedMNIST}
\end{subfigure}
\begin{subfigure}[b]{0.19\textwidth}
    \includegraphics[width=\textwidth]{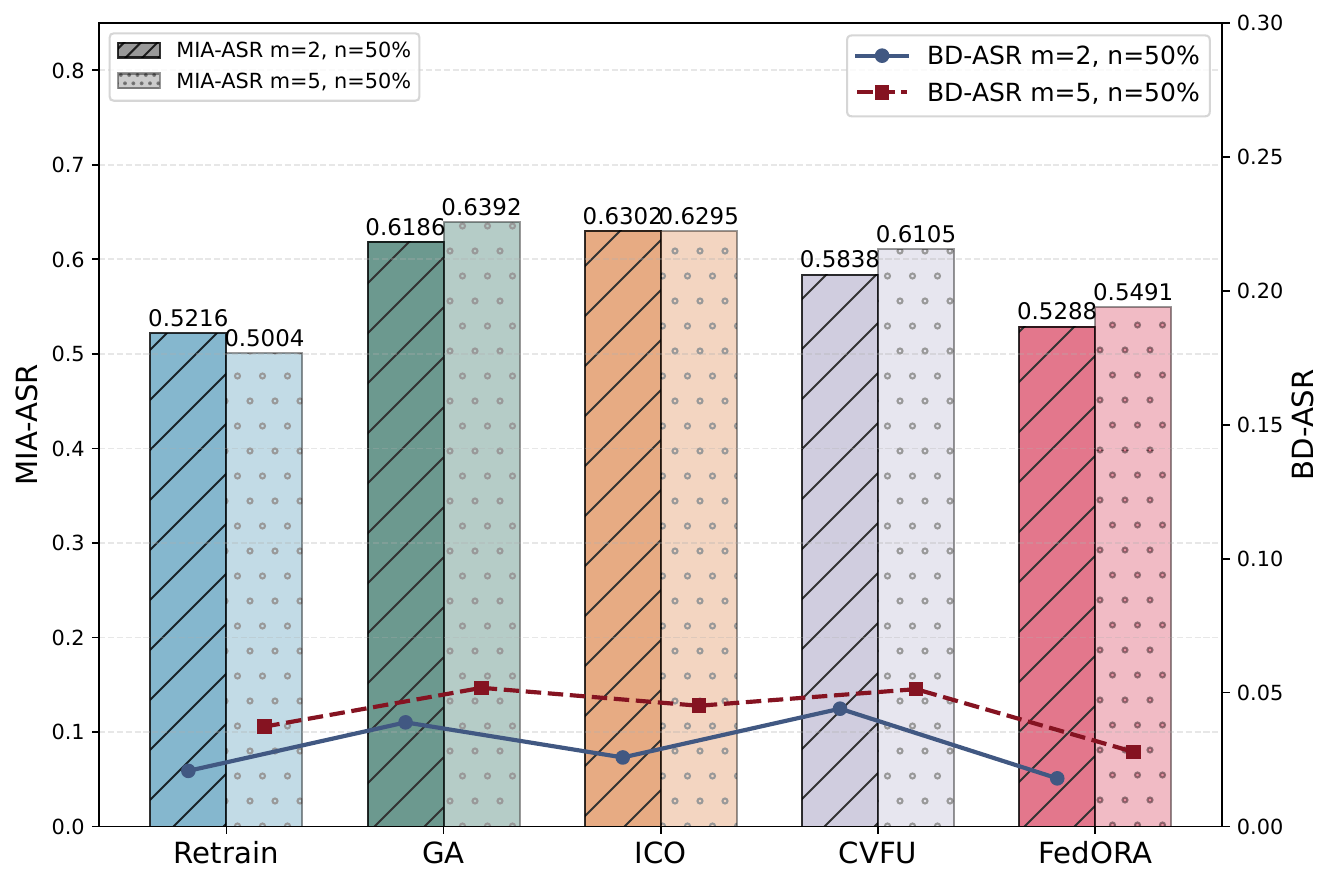}
    \caption{CIFAR-10}
\end{subfigure}
\begin{subfigure}[b]{0.19\textwidth}
    \includegraphics[width=\textwidth]{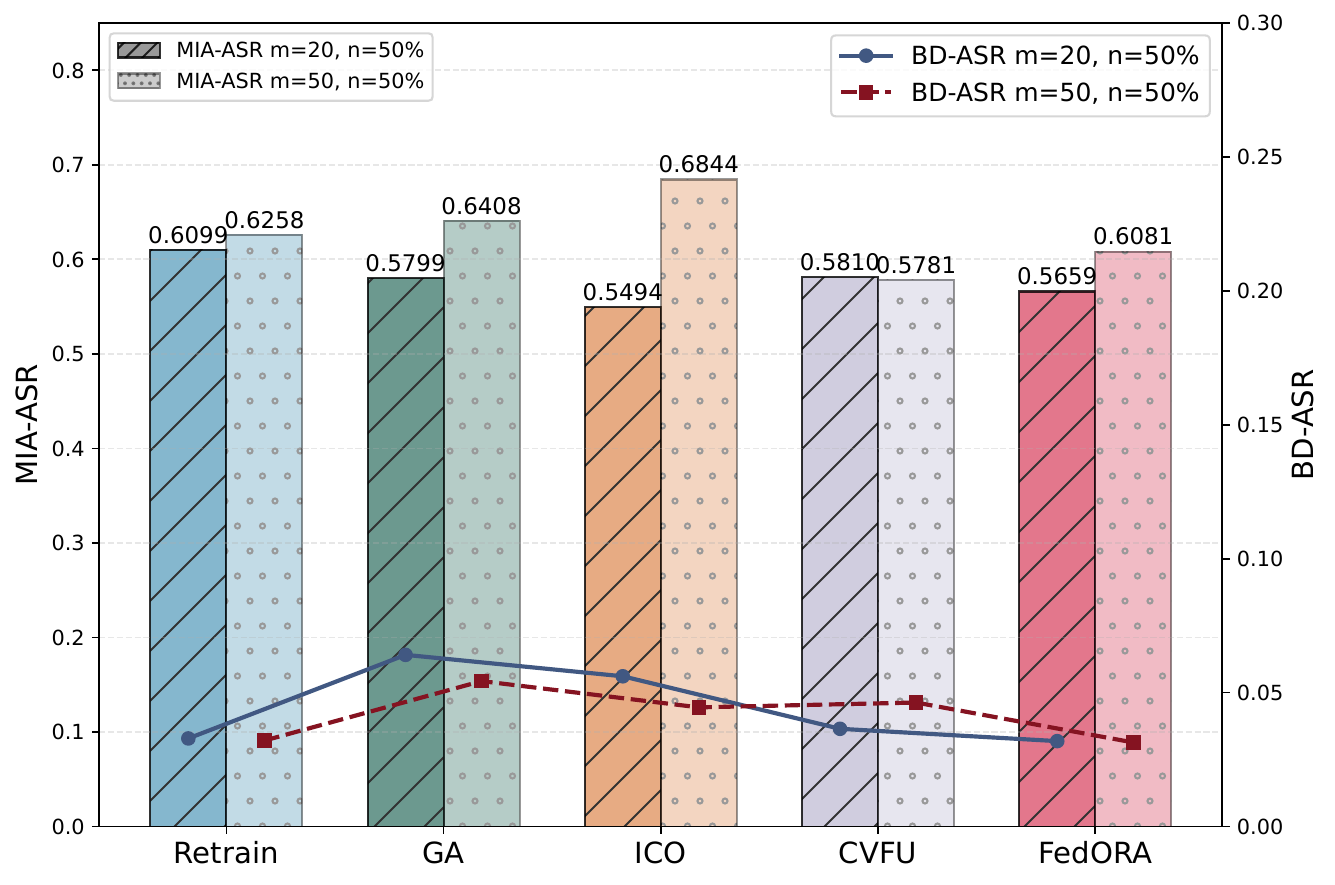}
    \caption{CIFAR-100}
\end{subfigure}
\begin{subfigure}[b]{0.19\textwidth}
    \includegraphics[width=\textwidth]{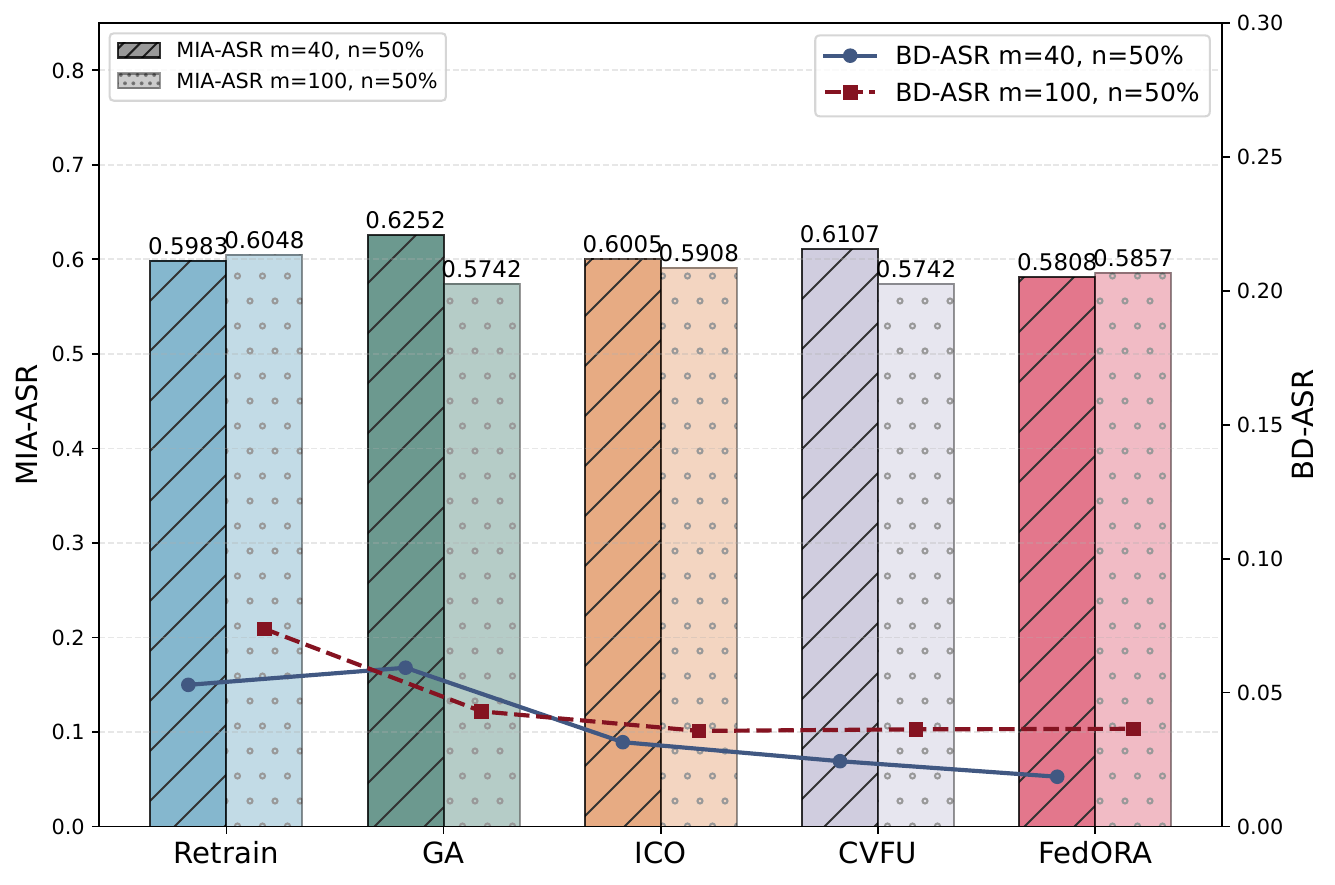}
    \caption{Tiny-ImageNet}
\end{subfigure}
\caption{{MIA-ASR and BD-ASR comparison across different sample unlearning settings.}}
\label{fig:mia_bd_comparison_sample}
\end{figure*}

\subsection{Evaluation}
\subsubsection{Complexity and Runtime Analysis}
Based on the complexity analysis in \cref{tab:complexity_comparison}, we present the computational and communication complexity across different VFL unlearning methods, where $K$ denotes the number of training iterations, $d$ represents the total model parameters, $d_h$ is the embedding dimension, and $B$ is the batch size.
From a computational perspective, GA achieves the lowest complexity by only processing the unlearn dataset $|D_u|$ over $K$ iterations, while FedORA maintains comparable efficiency when the batch selection ratio $\delta < 1$ is small. Retrain exhibits significantly higher computational cost due to reprocessing the entire remaining dataset over $K$ iterations, which is typically much larger than the unlearn set ($|D_r| \gg |D_u|$). ICO's complexity combines an initial unlearn phase $O(|D_u| d)$ with a refinement phase $O(|D_r| \nu K d)$, where the refinement factor $\nu < 1$ provides efficiency gains over full retraining. CVFU requires additional confidence matrix computation represented by the factor $\upsilon$, adding overhead to the base remaining dataset processing cost.
From a communication perspective, the relative efficiency rankings remain similar to computational costs, with GA being most efficient and Retrain being most expensive. FedORA reduces communication overhead through sampling with ratio $\delta$, making it more efficient than Retrain. ICO reduces refinement communication with factor $\nu$. CVFU incurs additional communication overhead $\upsilon$ for confidence score exchanges beyond standard embedding transmissions.

Based on the running time analysis in \cref{tab:running time}, we observe consistent performance patterns across datasets of varying complexity.
GA demonstrates the most efficient execution time across all datasets, as processing only the small unlearn dataset minimizes computational overhead.
FedORA achieves the second-fastest runtime among all methods, requiring only slightly more time than GA despite using dual-variable optimization.
ICO presents moderate performance due to computational overhead from its two-phase design, though it shows improved relative efficiency on complex datasets.
Retrain and CVFU exhibit the highest computational costs, particularly on larger datasets. Retrain shows significant overhead due to reprocessing the entire remaining dataset, while CVFU demonstrates similar costs due to confidence matrix computations across the entire dataset.

\subsubsection{Utility Preservation}
Based on the test accuracy comparison in \cref{tab:ta_sample} and \cref{tab:ta_label}, we observe distinct utility preservation patterns across sample and label unlearning scenarios.
In sample unlearning scenarios, on the tabular Income dataset, FedORA maintains performance closely matching Retrain, while GA, ICO, and CVFU show moderate accuracy decline compared to Retrain, achieving only around 80\%. Moreover, unlearning 25\% samples from two classes achieves slightly higher test accuracy than unlearning 50\% samples from a single class.
For image datasets, performance patterns become more evident with increasing complexity. On MedMNIST, FedORA again approaches Retrain's performance, while other methods show some decline. As datasets become larger and more complex, all five unlearning methods experience accuracy degradation. On MedMNIST and CIFAR-10, both Retrain and FedORA maintain accuracy above 80\%, while on CIFAR-100 and Tiny-ImageNet, performance drops below 70\%. Moreover, the performance gap between FedORA and other methods becomes more apparent. GA, ICO, and CVFU exhibit substantial accuracy drops, while FedORA maintains stability close to Retrain. Additionally, increasing the number of unlearning classes worsens performance, with the accuracy gap between different unlearning settings becoming more substantial on complex datasets.
Label unlearning presents greater challenges than sample unlearning across all methods and datasets. When unlearning the same number of classes but with 100\% sample removal, all methods experience more severe accuracy decline compared to unlearning 50\% of samples. When unlearning more classes, accuracy also shows further deterioration. However, FedORA maintains its advantage, closely matching Retrain's performance and even slightly surpassing Retrain on CIFAR-100 and Tiny-ImageNet.

\begin{figure*}[t]
\centering
\begin{subfigure}[b]{0.24\textwidth}
    \includegraphics[width=\textwidth]{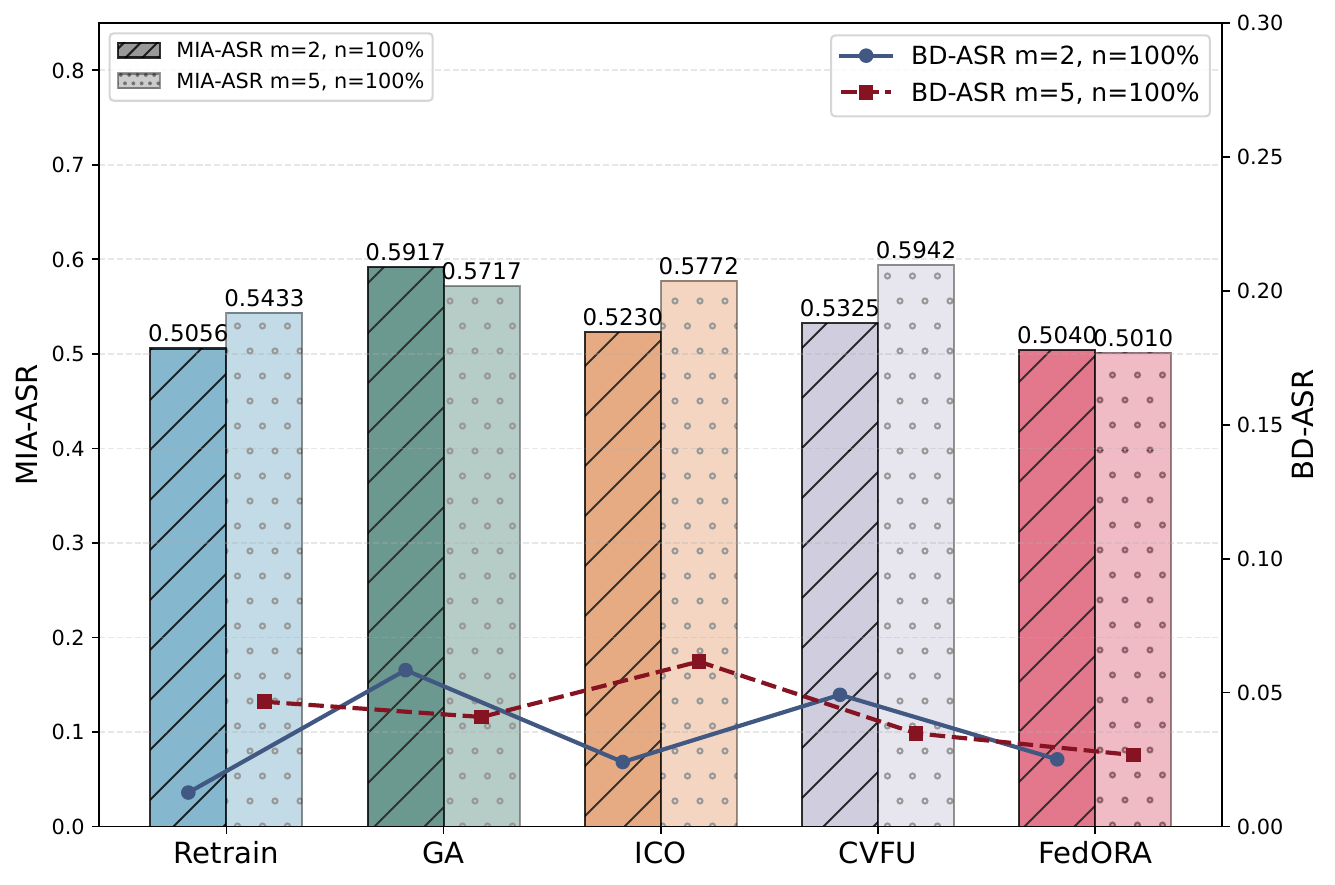}
    \caption{MedMNIST}
\end{subfigure}
\begin{subfigure}[b]{0.24\textwidth}
    \includegraphics[width=\textwidth]{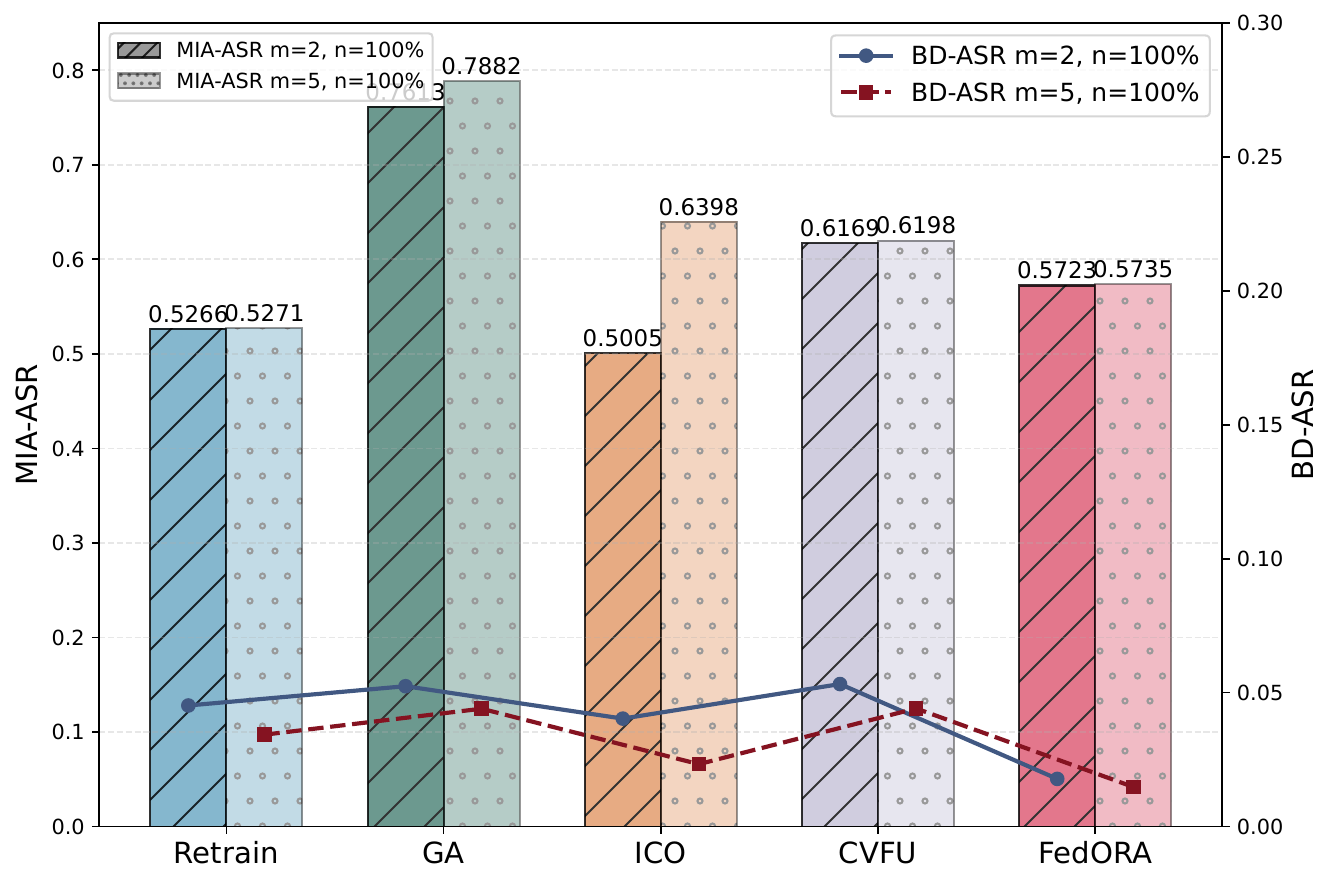}
    \caption{CIFAR-10}
\end{subfigure}
\begin{subfigure}[b]{0.24\textwidth}
    \includegraphics[width=\textwidth]{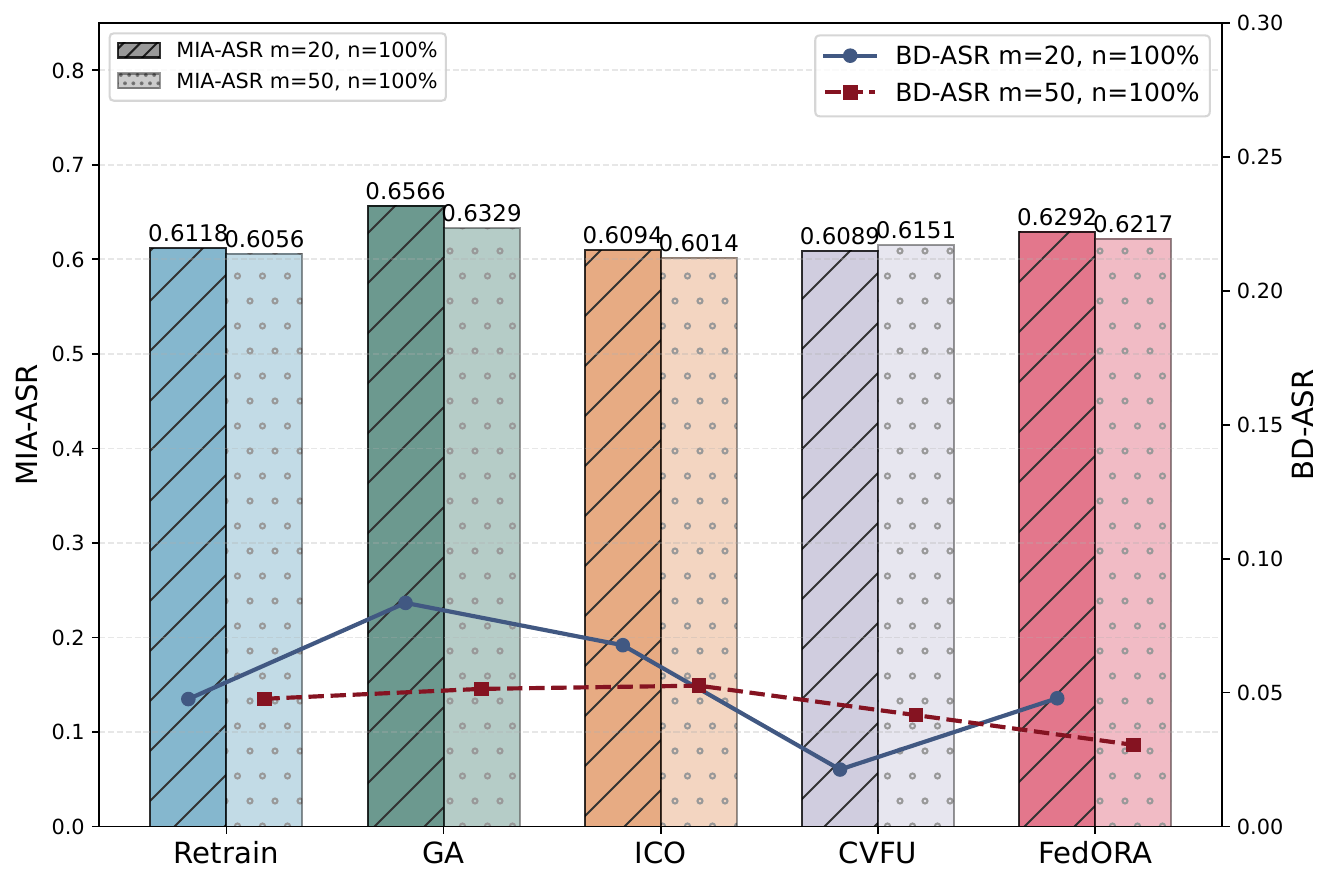}
    \caption{CIFAR-100}
\end{subfigure}
\begin{subfigure}[b]{0.24\textwidth}
    \includegraphics[width=\textwidth]{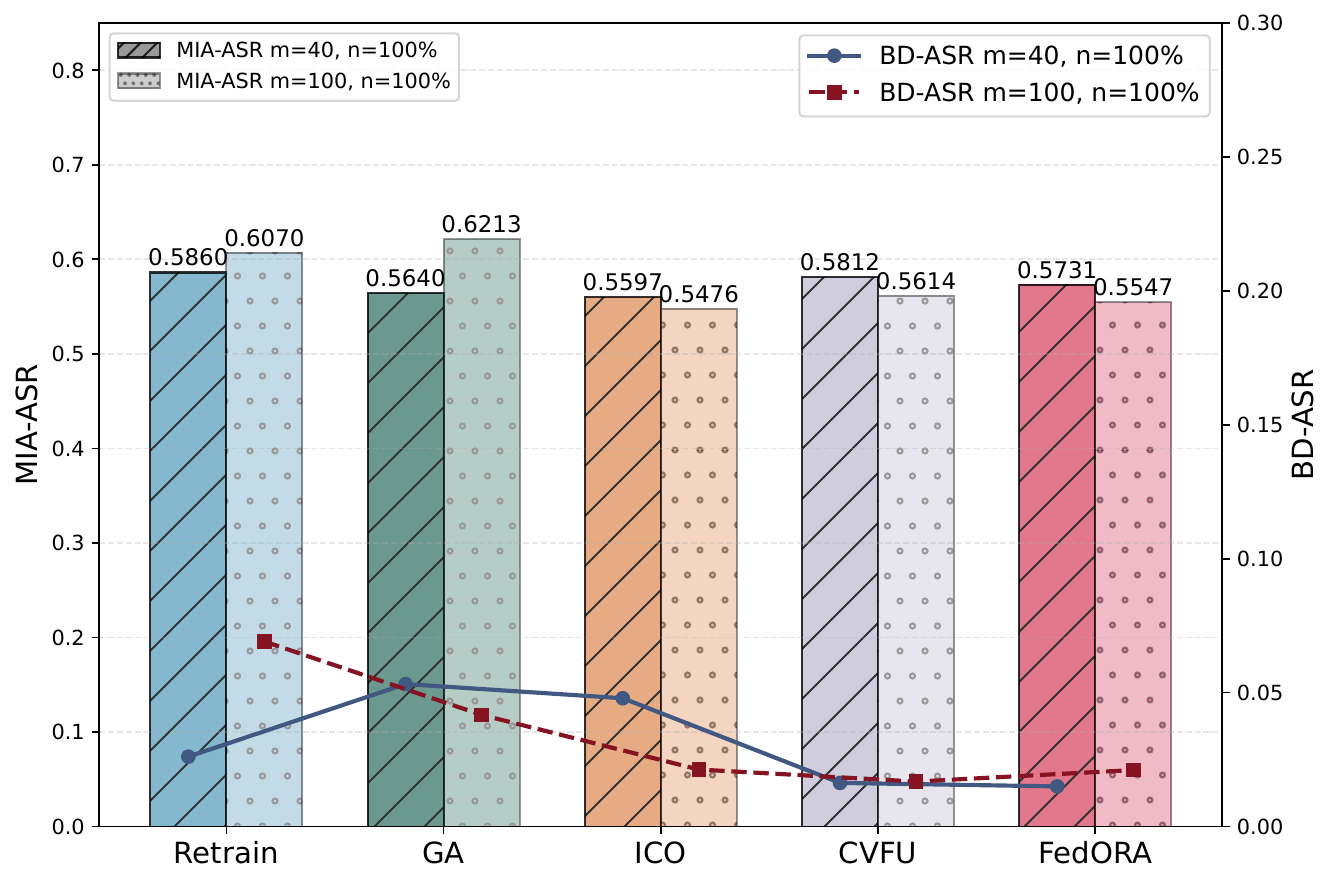}
    \caption{Tiny-ImageNet}
\end{subfigure}
\caption{{MIA-ASR and BD-ASR comparison across different label unlearning settings.}}
\label{fig:mia_bd_comparison_label}
\end{figure*}

\begin{table}[t]
\centering
\caption{{Performance comparison of different methods under IID and non-IID settings on CIFAR-10}}
\label{tab:noniid}
\begin{tabular}{llcccc}
\toprule
\multirow{2}{*}{Method} & \multirow{2}{*}{Setting} & Test & Unlearning & \multirow{2}{*}{MIA-ASR} & \multirow{2}{*}{BD-ASR} \\
& & Acc. & Acc. & & \\
\midrule
\multirow{2}{*}{Retrain} & IID & 0.8155 & 0.3028 & 0.5216 & 0.0208 \\
& Non-IID & 0.7357 & 0.2754 & 0.5579 & 0.0161 \\
\midrule
\multirow{2}{*}{GA} & IID & 0.6841 & 0.2830 & 0.6186 & 0.0389 \\
& Non-IID & 0.6342 & 0.3396 & 0.6379 & 0.0448 \\
\midrule
\multirow{2}{*}{ICO} & IID & 0.8079 & 0.3666 & 0.6302 & 0.0258 \\
& Non-IID & 0.7145 & 0.3507 & 0.5903 & 0.0234 \\
\midrule
\multirow{2}{*}{CVFU} & IID & 0.7848 & 0.3496 & 0.5838 & 0.0440 \\
& Non-IID & 0.7013 & 0.3922 & 0.6308 & 0.0534 \\
\midrule
\multirow{2}{*}{FedORA} & IID & 0.8165 & 0.2678 & 0.5288 & 0.0180 \\
& Non-IID & 0.7324 & 0.2620 & 0.5750 & 0.0225 \\
\bottomrule
\end{tabular}
\end{table}

\subsubsection{Unlearning Effectiveness}
We evaluate the unlearning effectiveness for five unlearning methods shown in \cref{fig:ua_sample} and \cref{fig:ua_label}.
We first analyze the sample unlearning scenario in \cref{fig:ua_sample}. On the Income dataset, all five methods exhibit higher unlearning accuracy relative to other datasets because Income is a binary classification dataset where random guessing has a higher probability of being correct, resulting in relatively higher unlearning accuracy. However, our FedORA achieves relatively lower unlearning accuracy compared to other methods. On image datasets, FedORA consistently demonstrates superior unlearning effectiveness, although other methods also achieve low unlearning accuracy. Additionally, we observe that GA exhibits greater volatility. Moreover, FedORA's advantage becomes more prominent on complex datasets. On CIFAR-100, Retrain's unlearning accuracy remains above 59\%. While GA, ICO, and CVFU achieve lower unlearning accuracy than Retrain, FedORA reaches below 35\%, demonstrating better unlearning effectiveness.
For the label unlearning scenario in \cref{fig:ua_label}, we find that all methods achieve lower unlearning accuracy compared to their corresponding sample unlearning results. This is reasonable because when unlearning partial samples from a class, the remaining samples still influence model performance, causing sample unlearning to have higher unlearning accuracy. On MedMNIST with two classes unlearned, Retrain achieves unlearning accuracy around 13\%, ICO and CVFU reach similar performance, GA performs worse with unlearning accuracy at 25.77\%, while FedORA demonstrates the best unlearning effectiveness with accuracy at 8.61\%, below 10\%. In other settings, FedORA's unlearning effectiveness is also nearly optimal.

\begin{table}[t]
\centering
\caption{{Ablation study of FedORA on CIFAR-10}}
\label{tab:ablation}
\begin{tabular}{lccccc}
\toprule
\multirow{2}{*}{} & Test & Unlearning & \multirow{2}{*}{MIA-ASR} & \multirow{2}{*}{BD-ASR} & Time \\
& Acc. & Acc. &  &  & (s) \\
\midrule
FedORA & 0.8165 & 0.2678 & 0.5288 & 0.0180 & 2.48 \\
w/o UL & 0.8351 & 0.8610 & 0.8448 & 0.9855 & 1.90 \\
w/o AB & 0.8235 & 0.2950 & 0.5344 & 0.0352 & 6.94 \\
w/o AS & 0.8078 & 0.3014 & 0.6643 & 0.0389 & 2.77 \\
\bottomrule
\end{tabular}\\
\raggedright\footnotesize{* UL: unlearning loss; AB: asymmetric batch design; AS: adaptive step sizes. w/o denotes ``without".}
\end{table}

\subsubsection{Attack Resilience}
We analyze the attack resilience of five methods against membership inference attacks and backdoor attacks on different datasets to assess unlearning performance, as shown in \cref{fig:mia_bd_comparison_sample} and \cref{fig:mia_bd_comparison_label}.
Firstly, we analyze MIA-ASR and BD-ASR under sample unlearning. On the Income dataset, Retrain, ICO, CVFU, and FedORA all achieve MIA-ASR close to 50\%, demonstrating effective unlearning, while GA performs slightly worse with MIA-ASR around 60\%. These methods achieve BD-ASR below 10\%, indicating that backdoor attacks are nearly ineffective under these unlearning approaches. For the image dataset MedMNIST, each method achieves MIA-ASR below 60\%, with ICO, CVFU, and FedORA ranging between 50\%-53\%. Additionally, the evaluated methods maintain BD-ASR below 5\%, confirming the effectiveness of unlearning. When unlearning more samples, each method's MIA-ASR and BD-ASR show slight increases compared to unlearning fewer samples. For more complex datasets CIFAR-10 and CIFAR-100, the methods exhibit increased MIA-ASR, while BD-ASR remains stable around 5\%. On the Tiny-ImageNet dataset, all approaches achieve MIA-ASR around 60\%, with Retrain and GA showing slightly higher BD-ASR compared to ICO, CVFU, and FedORA.
Under label unlearning, all five methods achieve lower MIA-ASR on MedMNIST compared to other datasets. GA shows slightly higher instability compared to other methods. ICO, CVFU, and FedORA essentially approach Retrain's performance. Additionally, these approaches achieve BD-ASR below 10\%, demonstrating effective unlearning.

\begin{figure}[t]
    \centering
        \begin{subfigure}[b]{0.49\linewidth}
        \includegraphics[width=\linewidth]{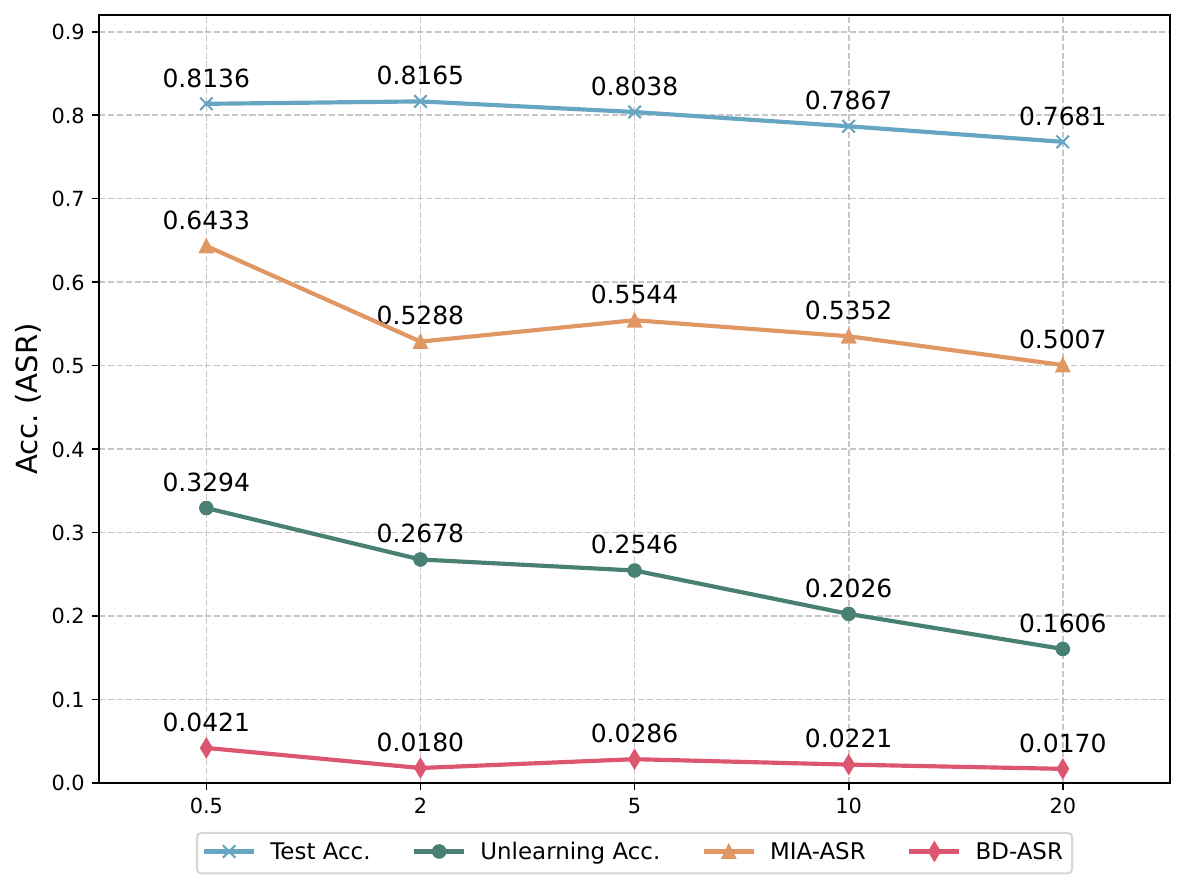}
        \caption{Forgetting weight $\omega$}
        \label{fig:pdhg_fweight_exp}
    \end{subfigure}
    \begin{subfigure}[b]{0.49\linewidth}
        \includegraphics[width=\linewidth]{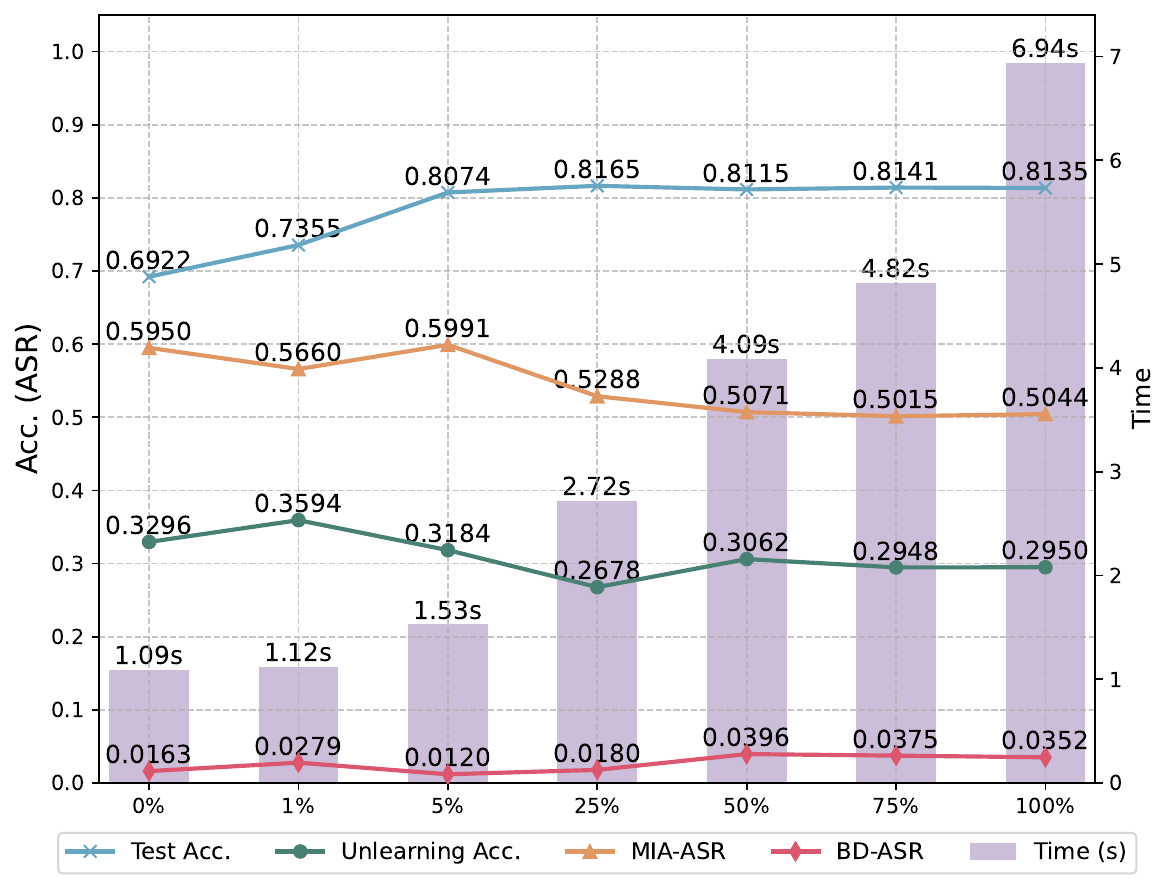}
        \caption{{Batch selection ratio $\delta$}}
        \label{fig:pdhg_batch_exp}
    \end{subfigure}
    \caption{Impact of forgetting weight and batch selection ratio.}
    \label{fig:sensitivity}
\end{figure}
\subsection{Analysis}
In this part, we conduct ablation experiments on the CIFAR-10 dataset under the unlearning setting of 2 classes with 50\% samples (m=2, n=50\%). We investigate FedORA's performance on IID and non-IID datasets. We explore the contribution of FedORA's different innovations on model performance, such as unlearning loss, asymmetric batch design, and adaptive step sizes. Furthermore, we conduct sensitivity analysis on the forgetting weight in the unlearning loss and the batch selection ratio in the asymmetric batch design to examine their impact on model performance.
\subsubsection{Performance on Non-IID Dataset}
VFL scenarios often involve non-IID data distributions across participating parties, making it crucial to evaluate unlearning methods under non-IID conditions. To simulate this setting, we introduce noise to images in one passive party while keeping the other party's data unchanged, creating data distribution heterogeneity.
As shown in \cref{tab:noniid}, we compare different methods on IID and non-IID datasets. Under non-IID conditions, all methods experience notable performance decline, with test accuracy dropping approximately 8\% across different approaches. Unlearning accuracy, MIA-ASR, and BD-ASR show slight fluctuations.
Retrain and ICO have slightly better unlearning performance under non-IID conditions, while other methods show slight declines in unlearning effectiveness, but the decline is limited. All methods maintain their unlearning capabilities under non-IID data conditions.

\subsubsection{Ablation Study}
We conduct ablation experiments to evaluate the contribution of each key innovation in FedORA's design. \cref{tab:ablation} presents the performance comparison when removing different components.
FedORA without unlearning loss (w/o UL) causes the most significant drop in unlearning effectiveness. Unlearning accuracy rises dramatically, showing poor unlearning ability as the model still retains knowledge of target samples. The clear increase in both MIA-ASR and BD-ASR further confirms reduced unlearning effectiveness, indicating that the model fails to unlearn the target data.
We can observe that the asymmetric batch design mainly affects training efficiency rather than unlearning quality, through FedORA without asymmetric batch design (w/o AB). While performance metrics stay relatively stable, training time increases significantly, showing the design's important role in achieving practical efficiency.
Using fixed step sizes instead of adaptive ones (w/o AS) leads to moderate decline across all areas, with lower test accuracy and weaker unlearning effectiveness.
Each component addresses specific challenges in the VFL unlearning process: the unlearning loss ensures effective forgetting, the asymmetric batch design enables fast training while keeping good performance, and adaptive step sizes contribute to better overall optimization performance in our primal-dual framework.

\subsubsection{Impact of Forgetting Weight}
We provide sensitivity analysis of the forgetting weight $\omega$ in \cref{fig:pdhg_fweight_exp}. As the forgetting weight increases from 0.5 to 20, test accuracy gradually decreases, indicating that higher forgetting weights compromise the model's performance. Meanwhile, unlearning accuracy drops from 32.94\% to 16.06\%, demonstrating more effective unlearning of target data. MIA-ASR also decreases as the forgetting weight increases, indicating improved unlearning effectiveness. BD-ASR generally maintains low values across all settings, showing consistent removal of backdoor patterns.
The forgetting weight demonstrates a trade-off between utility preservation and unlearning effectiveness. At $\omega = 2$, there is an optimal balance where the model achieves competitive test accuracy while maintaining effective unlearning performance.

\subsubsection{Impact of Batch Selection Ratio}
We analyze the impact of the batch selection ratio $\delta$ on FedORA's performance and efficiency, shown in \cref{fig:pdhg_batch_exp}. When the unlearning setting involves 2 classes with 50\% of the samples in CIFAR-10, equivalent to unlearning 10\% of the entire dataset, test accuracy improves dramatically from 69.22\% at $\delta=0\%$ to 80.74\% at $\delta=5\%$, reaching 81.65\% at $\delta=25\%$ before plateauing, demonstrating that processing only 5\% of the remaining data provides sufficient information for effective model updates. Unlearning accuracy decreases to 26.78\% at $\delta=25\%$ while MIA-ASR stabilizes around 50\% and BD-ASR remains consistently low. Importantly, reducing $\delta$ from 100\% to 5\% achieves a $4.5\times$ speedup.
We further evaluate FedORA under more extreme unlearning scenarios with fewer samples, as shown in \cref{tab:extreme}. When unlearning 50 samples, test accuracy improves from 74.15\% at $\delta=0\%$ to 80.54\% at $\delta=1\%$, while unlearning accuracy, MIA-ASR, and BD-ASR collectively demonstrate effective unlearning performance. 
In the most extreme case of unlearning only 1 sample, unlearning accuracy and BD-ASR drop to 0\%, and MIA-ASR stabilizes at 50\% approaching random guessing, confirming complete removal of the sample's influence. Notably, test accuracy shows little change between $\delta=0\%$ and $\delta=1\%$, indicating that processing only 1\% of remaining data suffices for maintaining model utility. FedORA leverages asymmetric batching with an adjustable selection ratio $\delta$ to reduce communication overhead, enabling efficient unlearning while maintaining model utility in large-scale VFL scenarios.

\subsubsection{Performance under Privacy-Preserving VFL}
Privacy-preserving mechanisms such as differential privacy (DP) are beneficial to VFL deployments. Accordingly, we analyze FedORA's compatibility with these techniques.
\cref{tab:gaussian_noise} presents FedORA's performance under varying levels of Gaussian noise injection. As noise decreases from $5.0 \times 10^{-4}$ to $0.1 \times 10^{-4}$, test accuracy improves from 56.72\% to 79.48\%, reflecting the privacy-utility trade-off. Despite utility degradation under stronger noise, FedORA maintains effective unlearning across all noise levels, with unlearning accuracy consistently below 30\%, MIA-ASR stabilizing around 54-57\%, and BD-ASR remaining below 10\%. These results demonstrate FedORA's robustness and compatibility with privacy-preserving mechanisms, enabling its deployment in privacy-critical VFL.

\begin{table}[t]
\centering
\caption{{FedORA under extreme unlearning scenarios}}
\label{tab:extreme}
\begin{tabular}{cccccc}
\toprule
No. of &$\delta$  & Test & Unlearning &  \multirow{2}{*}{MIA-ASR} &\multirow{2}{*}{BD-ASR}  \\
Samples  & (\%) & Acc. & Acc. &&   \\
\midrule
\multirow{2}{*}{50} & 0 & 0.7415 & 0.1000 & 0.5628 & 0.0091  \\
 & 1 & 0.8054 & 0.1400 & 0.5533 & 0.0095  \\
\midrule
\multirow{2}{*}{1} & 0 & 0.7916 & 0.0000 & 0.5000 & 0.0000  \\
 & 1 & 0.8117 & 0.0000 & 0.5000 & 0.0000  \\
\bottomrule
\end{tabular}
\end{table}

\begin{table}[t]
\centering
\caption{{FedORA under different levels of Gaussian noise}}
\label{tab:gaussian_noise}
\begin{tabular}{ccccc}
\toprule
Gaussian Noise  & Test & Unlearning & \multirow{2}{*}{MIA-ASR} & \multirow{2}{*}{BD-ASR} \\
($\times 10^{-4}$) & Acc. & Acc. &  &   \\
\midrule
5.0 & 0.5672 & 0.1632 & 0.5467 & 0.0453  \\
1.0 & 0.6725 & 0.2062 & 0.5509 & 0.0858  \\
0.5 & 0.7237 & 0.2308 & 0.5719 & 0.0547  \\
0.1 & 0.7948 & 0.2766 & 0.5458 & 0.0296  \\
\bottomrule
\end{tabular}
\end{table}

\section{Limitations and Future Work}
\textbf{Unlearning Scope.} FedORA focuses on sample unlearning that removes partial samples from specific classes while preserving other samples, and label unlearning that involves complete removal of entire classes along with all associated samples.
Our approach focuses on label deletion rather than modification or correction. However, label modification and correction are common in real-world applications, requiring simultaneous forgetting of old labels and learning of new labels. 
In the future work, we consider extending our framework to support modification and correction scenarios, requiring additional considerations for dual variable updates and constraint formulations.

\textbf{Model Scale and Scalability.} While our work demonstrates effective unlearning in VFL using conventional neural network architectures, extension to large language models (LLM) remains unexplored due to computational and architectural constraints.
Recent advances~\cite{lin2024splitlora,lin2025hsplitlora,han2024parameter} demonstrate the potential for applying federated learning paradigms to large language models through parameter-efficient fine-tuning techniques. However, adapting our primal-dual unlearning framework to LLM would require addressing additional challenges including parameter-efficient fine-tuning integration and specialized handling of transformer architectures in VFL settings. SplitLoRA~\cite{lin2024splitlora} and HSplitLoRA~\cite{lin2025hsplitlora} provide valuable insights into how model splitting and LoRA can be combined for efficient LLM training, which could potentially be extended to support unlearning scenarios. In the future work, we can integrate our primal-dual framework with parameter-efficient techniques like LoRA for large-scale model unlearning.

\textbf{VFL-Specific Attacks.} FedORA demonstrates robustness in VFL deployments with DP, achieving stable optimization and effective unlearning across noise levels. Future work could explore more VFL-specific attacks. Such research could leverage passive party feature inference or reconstruction methods to validate unlearning effectiveness, or examine how unlearning processes affect vulnerability to these attacks in vertical federated settings.

\section{Conclusion}
In summary, we address the significant challenge of sample and label unlearning in vertical federated learning. We introduce FedORA, an approach that formulates unlearning as a constrained optimization problem solved through a primal-dual framework. 
In our approach, the proposed loss function encourages uncertainty in classification for unlearning data, rather than forcing misclassification. This helps to mitigate the excessive forgetting often observed in gradient ascent methods, while more effectively removing the influence of the target data. Furthermore, through Lagrange duality, the resulting dual structure naturally provides a certificate of both feasibility and optimality for the unlearning process.
To enhance stability, we integrate an adaptive step size mechanism. FedORA uses an asymmetric batch processing strategy, processing full batches of unlearning data for complete unlearning and partial batches of remaining data to reduce computational overhead while preserving model utility.
We provide theoretical analysis proving that the model difference between FedORA and Train-from-scratch is bounded, establishing guarantees for the unlearning effectiveness of FedORA.
Extensive experiments on tabular and image datasets across multiple unlearning scenarios demonstrate that FedORA effectively balances unlearning effectiveness with model utility preservation while significantly improving computational and communication efficiency.

% \section*{Acknowledgment}
% This research is supported by the National Research Foundation, Singapore and Infocomm Media Development Authority under its Trust Tech Funding Initiative, the Ministry of Education Academic Research Fund MOE Tier 2 Grant MOE-T2EP20224-0009, National Natural Science Foundation of China Grant 62472460, Guangdong Basic and Applied Basic Research Foundation Grants 2024A1515010161 and 2023A1515012982, and Young Outstanding Award under the Zhujiang Talent Plan of Guangdong Province. Any opinions, findings and conclusions or recommendations expressed in this material are those of the author(s) and do not reflect the views of National Research Foundation, Singapore and Infocomm Media Development Authority.

\ifCLASSOPTIONcaptionsoff
  \newpage
\fi

\bibliographystyle{IEEEtran}
\bibliography{mybibliography}

\appendices

\section{Proof of Theorem \ref{theo}}
\label{Proof of Theorem}
According to \cref{eq:dual update} and \cref{eq:primal update}, the primal update of FedORA follows: 
\begin{equation}
\begin{aligned}
        \Theta^{k+1} = & \Theta^k - \tau (\nabla \mathcal{L}_r(\Theta^k) + \epsilon_r^k - \nabla \mathcal{L}_u(\Theta^k) \circ \Omega^{k+1} \\
        & + \rho(\Theta^k - \Theta_{init})),
\end{aligned}
\end{equation}
where $\epsilon_r^k$ represents the mini-batch noise satisfying $\|\epsilon_r^k\|_2 \leq \frac{\sigma_r}{\sqrt{|\mathcal{B}_r|}}$ under Assumption \ref{assump:minibatch}. 
The Train-from-scratch update follows: $\bar{\Theta}^{k+1} = \bar{\Theta}^k - \tau \nabla \mathcal{L}_r(\bar{\Theta}^k)$.
For simplicity, we do not consider the regularization term and assume that the step size for the primal update in FedORA and the step size for retraining are the same, denoted as $\tau$.
We define $e^k = \Theta^k - \bar{\Theta}^k$ to measure the model difference between FedORA and Train-from-scratch. From the update equations of FedORA and Train-from-scratch, we obtain:
\begin{equation}
\begin{aligned}
    e^{k+1} =& e^k - \tau (\nabla \mathcal{L}_r(\Theta^k) - \nabla \mathcal{L}_r(\bar{\Theta}^k) \\
        & - \nabla \mathcal{L}_u(\Theta^k) \circ \Omega^{k+1} + \epsilon_r^k).
\end{aligned}
\end{equation}
By applying the norm expansion formula, we can expand:
\begin{equation}
    \begin{aligned}
        \|e^{k+1}\|^2_2 =& \|e^k\|^2_2 + \tau^2 \|g^k\|^2_2 - 2\tau \langle e^k, \epsilon_r^k \rangle \\
         &- 2\tau \langle e^k, \nabla \mathcal{L}_r(\Theta^k) - \nabla \mathcal{L}_r(\bar{\Theta}^k) \rangle \\
        & + 2\tau \langle e^k, \nabla \mathcal{L}_u(\Theta^k) \circ \Omega^{k+1} \rangle ,
    \end{aligned}
\end{equation}
where $g^k = \nabla \mathcal{L}_r(\Theta^k) - \nabla \mathcal{L}_r(\bar{\Theta}^k) + \epsilon_r^k - \nabla \mathcal{L}_u(\Theta^k) \circ \Omega^{k+1}$.

To analyze the boundedness of $\|g^k\|^2_2$, we first apply Assumption \ref{assump:convex} which gives us:
\begin{equation}
    \|\nabla \mathcal{L}_r(\Theta^k) - \nabla \mathcal{L}_r(\bar{\Theta}^k)\|_2 \leq L \|\Theta^k - \bar{\Theta}^k\|_2 = L \|e^k\|_2.
\end{equation}
The mini-batch noise term is bounded by Assumption \ref{assump:minibatch}: $\|\epsilon_r^k\|_2 \leq \frac{\sigma_r}{\sqrt{|\mathcal{B}_r|}}$.
Under the convergence theory of the primal-dual algorithms, appropriate step size selection ensures the boundedness of the dual variable sequence, such that $\|\Omega^{k+1}\|_2 \leq \Omega_{max}$, where $\Omega_{max}$ is a bounded constant. The unlearning term is bounded by Assumption \ref{assump:gradient} and the boundedness of dual variables:
\begin{equation}
\begin{aligned}
     \|\nabla \mathcal{L}_u(\Theta^k) \circ \Omega^{k+1}\|_2 
     &\leq \|\nabla \mathcal{L}_u(\Theta^k)\|_{\infty} \|\Omega^{k+1}\|_2 \\
    &\leq \|\nabla \mathcal{L}_u(\Theta^k)\|_2 \|\Omega^{k+1}\|_2 \\
    &\leq G \Omega_{max}.
\end{aligned}
\end{equation}
By the triangle inequality, we can obtain:
\begin{equation}
    \|g^k\|_2 \leq L \|e^k\|_2 + \frac{\sigma_r}{\sqrt{|\mathcal{B}_r|}} + G \Omega_{max}.
\end{equation}
Based on Assumption \ref{assump:convex}, we can derive:
\begin{equation}
    -2\tau \langle e^k, \nabla \mathcal{L}_r(\Theta^k) - \nabla \mathcal{L}_r(\bar{\Theta}^k) \rangle \leq -2\tau \mu \|e^k\|^2_2.
\end{equation}
For the mini-batch term, we have:
\begin{equation}
    |\langle e^k, \epsilon_r^k \rangle| \leq \|e^k\|_2 \|\epsilon_r^k\|_2 \leq \frac{\sigma_r}{\sqrt{|\mathcal{B}_r|}} \|e^k\|_2 .
\end{equation}
The unlearning term satisfies under Assumption \ref{assump:gradient}:
\begin{equation}
\begin{aligned}
        |\langle e^k, \nabla \mathcal{L}_u(\Theta^k) \circ \Omega^{k+1} \rangle| &\leq \|e^k\|_2 \|\nabla \mathcal{L}_u(\Theta^k)\|_2 \|\Omega^{k+1}\|_2 \\
        &\leq G \|e^k\|_2 \|\Omega^{k+1}\|_2.
\end{aligned}
\end{equation}
Therefore, $ |\langle e^k, \nabla \mathcal{L}_u(\Theta^k) \circ \Omega^{k+1} \rangle| \leq G \Omega_{max} \|e^k\|_2$.
For the cross terms, we apply Young's inequality $2ab \leq a^2 + b^2$:
\begin{equation}
\begin{aligned}
        &2\tau \|e^k\|_2 \left(\frac{\sigma_r}{\sqrt{|\mathcal{B}_r|}} + G \Omega_{max}\right) \\ \leq & \tau \|e^k\|_2^2 + \tau \left(\frac{\sigma_r}{\sqrt{|\mathcal{B}_r|}} + G \Omega_{max}\right)^2.
\end{aligned}
\end{equation}

Substituting the bounds back into the expansion of $\|e^{k+1}\|^2_2$, we have:
\begin{equation}
\begin{aligned}
        \|e^{k+1}\|^2_2 
        \leq & \|e^k\|^2_2 - 2\tau \mu \|e^k\|^2_2 + \tau \|e^k\|^2_2 + \\
        &\tau \left(\frac{\sigma_r}{\sqrt{|\mathcal{B}_r|}} + G \Omega_{max}\right)^2 + \tau^2 \|g^k\|^2_2.
\end{aligned}
\end{equation}
For a sufficiently small step size $\tau$ such that $\tau \leq \min\left\{\frac{1}{2L}, \frac{\mu}{4L^2}\right\}$, the quadratic term $\tau^2 \|g^k\|^2_2$ becomes negligible compared to the linear terms. This gives us:
\begin{equation}
\begin{aligned}
        \|e^{k+1}\|^2_2 
        &\leq (1 - \tau \mu)\|e^k\|^2_2 + \tau \left(\frac{\sigma_r}{\sqrt{|\mathcal{B}_r|}} + G \Omega_{max}\right)^2.
\end{aligned}
\end{equation}
Taking the square root of both sides and applying the inequality $\sqrt{a^2 + b^2} \leq \sqrt{a^2} + \sqrt{b^2}$ for $a, b \geq 0$, we have:
\begin{equation}
    \|e^{k+1}\|_2 \leq \sqrt{1 - \tau \mu} \|e^k\|_2 + \sqrt{\tau} \left(\frac{\sigma_r}{\sqrt{|\mathcal{B}_r|}} + G \Omega_{max}\right).
\end{equation}
The above inequality represents a standard linear recursion of the form $a_{k+1} \leq A_1 a_k + A_2$, where $A_1 = \sqrt{1 - \tau \mu} < 1$ is the contraction factor and $A_2 = \sqrt{\tau} \left(\frac{\sigma_r}{\sqrt{|\mathcal{B}_r|}} + G \Omega_{max}\right)$ is the bias term. Since $\tau \mu > 0$, we have $A_1 < 1$, which ensures the convergence of the recursion. The solution to such a recursion is $ \|e^k\|_2 \leq A_1^k \|e^0\|_2 + \frac{A_2}{1-A_1}$.
Substituting the specific values:
\begin{equation}
\begin{aligned}
        \|\Theta^k - \bar{\Theta}^k\|_2 \leq & (\sqrt{1 - \tau \mu})^k \|\Theta^0 - \bar{\Theta}^0\|_2 \\
        + &\frac{\sqrt{\tau}}{1 - \sqrt{1-\tau\mu}} \left(\frac{\sigma_r}{\sqrt{|\mathcal{B}_r|}} + G \Omega_{max}\right).
\end{aligned}
\end{equation}
As $k \to \infty$:
\begin{equation}
    \lim_{k \to \infty} \|\Theta^k - \bar{\Theta}^k\|_2 \leq \frac{\sqrt{\tau}}{1 - \sqrt{1-\tau\mu}} \left(\frac{\sigma_r}{\sqrt{|\mathcal{B}_r|}} + G \Omega_{max}\right).
\end{equation}

\end{document}